\documentclass[fleqn,usenatbib]{mnras}

\usepackage[T1]{fontenc}

\DeclareRobustCommand{\VAN}[3]{#2}
\let\VANthebibliography\thebibliography
\def\thebibliography{\DeclareRobustCommand{\VAN}[3]{##3}\VANthebibliography}

\usepackage{graphicx}	
\usepackage{amsmath}	
\usepackage{amssymb}	
\usepackage[dvipsnames]{xcolor}
\usepackage{multirow}
\usepackage{delarray}
\usepackage{color}
\usepackage{here}
\usepackage{float}
\usepackage{tensor}
\usepackage{xspace}
\usepackage{orcidlink}
\usepackage[multiple]{footmisc}
\usepackage{newtxtext,newtxmath}
\defcitealias{guszejnov_isothermal_mhd}{Paper 0}
\defcitealias{starforge.methods}{Paper I}
\defcitealias{starforge_jets_imf}{Paper II}
\usepackage[normalem]{ulem}
\newcommand{\msun}{\mathrm{M}_{\rm \sun}}

\usepackage{graphicx}	
\usepackage{amsmath}	
\usepackage{amssymb}	

\title[Star formation with all feedback in concert]{The dynamics and outcome of star formation with jets, radiation, winds, and supernovae in concert}

\author[M. Y. Grudi\'{c} et al.]{
Michael Y. Grudi\'{c}\orcidlink{0000-0002-1655-5604}$^{1}$\thanks{mgrudic@carnegiescience.edu}\thanks{NASA Hubble Fellow},
D\'avid Guszejnov\orcidlink{0000-0001-5541-3150}$^{2}$\thanks{guszejnov@utexas.edu},
Stella S. R. Offner\orcidlink{0000-0003-1252-9916}$^{2}$,
Anna L. Rosen\orcidlink{0000-0003-4423-0660}$^{4}$,\newauthor
Aman N. Raju\orcidlink{0000-0001-9339-0789}$^{2}$, 
Claude-Andr{\'e} Faucher-Gigu{\`e}re\orcidlink{0000-0002-4900-6628}$^{3}$, and
Philip F. Hopkins\orcidlink{0000-0003-3729-1684}$^{5}$
\\
$^{1}$Carnegie Observatories, 813 Santa Barbara St, Pasadena, CA 91101, USA\\
$^{2}$Department of Astronomy, The University of Texas at Austin, TX 78712, USA \\
$^{3}$CIERA and Department of Physics and Astronomy, Northwestern University, 1800 Sherman Ave, Evanston, IL 60201, USA\\
$^{4}$Center for Astrophysics $|$ Harvard \& Smithsonian, 60 Garden St, Cambridge, MA 02138, USA \\
$^{5}$TAPIR, Mailcode 350-17, California Institute of Technology, Pasadena, CA 91125, USA \\
}

\date{Accepted XXX. Received YYY; in original form ZZZ}

\pubyear{2021}

\begin{document}
\label{firstpage}
\pagerange{\pageref{firstpage}--\pageref{lastpage}}
\maketitle

\begin{abstract}
We analyze the first giant molecular cloud (GMC) simulation to follow the formation of individual stars and their feedback from jets, radiation, winds, and supernovae, using the {\small STARFORGE} framework in the {\small GIZMO} code. We evolve the GMC for $\sim 9 \rm Myr$, from initial turbulent collapse to dispersal by feedback. Protostellar jets dominate feedback momentum initially, but radiation and winds cause cloud disruption at $\sim 8\%$ star formation efficiency (SFE), and the first supernova at $8.3 \rm Myr$ comes too late to influence star formation significantly. The per-freefall SFE is dynamic, accelerating from 0 to $\sim 18\%$ before dropping quickly to <1\%, but the estimate from YSO counts compresses it to a narrower range. The primary cluster forms hierarchically and condenses to a brief ($\sim 1\,\mathrm{Myr}$) compact ($\sim 1 \rm pc$) phase, but does not virialize before the cloud disperses, and the stars end as an unbound expanding association. The initial mass function resembles the Chabrier (2005) form with a high-mass slope $\alpha=-2$ and a maximum mass of $55 M_\odot$. Stellar accretion takes $\sim 400 \rm kyr$ on average, but $\gtrsim 1\rm Myr$ for $>10 M_\odot$ stars, so massive stars finish growing latest. The fraction of stars in multiples increases as a function of primary mass, as observed. Overall, the simulation much more closely resembles reality, compared to previous versions that neglected different feedback physics entirely. But more detailed comparison with synthetic observations will be needed to constrain the theoretical uncertainties.
\end{abstract}




\section{Introduction}
The basic story of how stars form has long been established: they form mainly in giant molecular clouds (GMCs) with masses $\sim 10^4 -10^7 M_\odot$ \citep{goldreich.kwan.gmcs,zuckerman.evans.gmcs, williams_mckee_1997}, due to the fragmentation and collapse of gravitationally-unstable cores \citep{Jeans_1902,larson_1969, rosen_2020_review}. New stars generally form in relative proximity to other young stars, i.e. in clusters \citep{clustering_lada, krumholz_2019_cluster_review}.  Yet many important details of star formation (SF) are not well understood, such as {\it why} stars have the particular masses that they do (i.e. the origin of the initial mass function, IMF), why and how they form in clusters, why they apparently form with such low efficiency, and why some are in multiple systems and others are not. 

Theoretical and computational models can offer insights into these questions, but it has proven challenging to produce a detailed model of SF realistic enough to reproduce the basic hallmarks of star formation. Star-by-star simulations of star cluster formation have progressed for more than two decades \citep{klessen00a,bate2003,Offner_2009_radiative_sim,Federrath_2010_sinks,Haugbolle_Padoan_isot_IMF,Cunningham_2018_feedback,mathew:2021.imf.jets}, but only recently have started simulating star formation over the spatial ($10+\rm pc$), temporal (several Myr) and GMC mass ($\gtrsim 10^4 M_\odot$) scales that are directly comparable to well-studied nearby star-forming GMCs and young star clusters \citep[e.g.][]{hillenbrand:1998.orion.imf, hsu:2012.orion.a.lowmass.imf, Evans_2014_MW_GMC_SFR, pokhrel:2020.gmc.sfe}, where important quantities such as star formation efficiency and the IMF are the most well-constrained, and massive ($\gtrsim 10 M_\odot$) stars can form.

In \citet{guszejnov_isothermal_mhd} (hereafter \citetalias{guszejnov_isothermal_mhd}) we ran a large suite of GMC simulations accounting for gravity and MHD turbulence with the {\small GIZMO} code's Meshless Finite-Mass (MFM) method \citep{hopkins_gizmo_mhd}. We found these models inevitably predicted excessively-high star formation efficiency and an extreme excess of massive stars in Milky Way-like conditions, implying that additional mechanisms are important for star formation, and in particular some form of feedback must moderate stellar accretion. In \citet{starforge.methods} (hereafter \citetalias{starforge.methods}) we introduced the more-advanced {\small STARFORGE}\footnote{\url{http://www.starforge.space}} framework for the {\small GIZMO} code, combining modules for gravity, N-body dynamics, MHD, radiative transfer, cooling and chemical physics, (proto-)stellar evolution, and feedback in the form of accretion- and fusion-powered radiation from stars and protostars, stellar winds, protostellar jets, and core-collapse supernovae. And in \citet{starforge_jets_imf} (hereafter \citetalias{starforge_jets_imf}) we used {\small STARFORGE} to re-run our GMC models with the addition of realistic ISM cooling/heating physics and protostellar jet feedback, finding that jet feedback in particular is crucial for moderating the growth of individual stars and recovering a realistic IMF, in agreement with other IMF studies with jet feedback \citep{hansen_lowmass_sf_feedback,krumholz_2012_orion_sims, Myers_2013_ORION_radiation_IMF, Federrath_2014_jets, mathew:2021.imf.jets}. 

Many GMCs simulated in \citetalias{starforge_jets_imf} still exhibited unrealistic phenomena, especially a high-mass excess in the IMF. Even $2\times 10^{4} M_\odot$ clouds could eventually form $>400 M_\odot$ stars through uninterrupted accretion, despite the extremely powerful jet feedback emanating from such stars in the model. Again, the natural explanation was missing feedback: very massive stars are near the \citet{eddington.limit} limit $L/L_\odot \sim 3.5 \times 10^4 M_{\star}/M_\odot$, around which radiation should drive instability or mass loss in overmassive stars \citep{stothers:1992.vms.stability,vink:2018.very.massive.stars}, and present a significant obstacle to the accretion of further gas \citep{larson:1971.masslimit, krumholz_2009_massive_sf,kuiper_2010_massive_sf,rosen_2016_massive_sf}.

In this paper we introduce the next phase of the {\small STARFORGE} project: the first GMC simulation run with the full physics package. This is the first numerical simulation of any kind to model the formation of a stellar cluster while tracking the formation, accretion, motion, evolution, and feedback of individual stars and protostars, with feedback from all major channels: protostellar jets, stellar winds, stellar radiation, and core-collapse supernovae. We aim to present a mile-wide, inch-deep picture of the outcome of the calculation, describing the overall sequence of GMC and star cluster evolution, and examining key SF outcomes: the star formation history, star cluster assembly, the impact of different feedback mechanisms,  star formation efficiency, the IMF, and stellar multiplicity. In future work, we will explore each of these subjects individually in much greater detail; here our goal is to survey the ensemble of key star formation predictions. Along the way we perform some basic comparisons of the simulation results to observations, to assess the overall fidelity of the full {\small STARFORGE} model.

This paper is organized as follows: in \S\ref{sec:methods} we describe the code, physics modules, and initial conditions used for the simulation. In \S\ref{sec:results} we present various results of the simulation, including overall global evolution of gas and stars, star formation efficiency, stellar accretion, the IMF, and stellar multiplicity. In \S\ref{sec:discussion} we compare our results to previous work and discuss various implications of the simulation's results. In \S\ref{sec:conclusion} we summarize our main findings. For the purposes of this paper, we refer to the entire population of stars formed in the same cloud as a ``cluster", making no distinction between bound and unbound members. When discussing the IMF, the ``IMF slope" we refer to is $\alpha$ such that $\mathrm{d}N/\mathrm{d} M_\star \propto M_{\star}^\alpha$, and $\alpha=-2.35$ corresponds to the canonical \citet{salpeter_slope} value. When making comparisons to the IMF and its statistics, we assume the \citet{chabrier_imf} form with an upper cutoff of $150M_\odot$.

\section{Methods}
\label{sec:methods}


We perform a 3D radiation MHD simulation of star cluster formation in a GMC with initial mass $M_{\rm 0}=2\times 10^4 M_\odot$ and radius $R_{\rm 0} = 10 \rm pc$ using the {\small STARFORGE} numerical framework implemented in the {\small GIZMO} code \citep{starforge.methods,hopkins2015_gizmo}, with all implemented feedback physics enabled: protostellar jets, radiation, winds, and core-collapse supernovae. 
The numerical implementation and tests of the {\small STARFORGE} modules are detailed fully in \citetalias{starforge.methods}, so here we only summarize them briefly.

\subsection{Magnetohydrodynamics}
\label{sec:mhd}
The simulation uses {\small GIZMO}'s mesh-free, quasi-Lagrangian Meshless Finite-Mass (MFM) magnetohydrodynamics (MHD) solver \citep{hopkins_gizmo_mhd}, and enable the \citet{Hopkins_2016_divb_cleaning} constrained-gradient scheme to control the $\nabla \cdot \mathbf{B}=0$ constraint to high precision. The fluid is initially discretized into equal-mass gas cells each containing the mass resolution $\Delta m = 10^{-3} M_\odot$, which move with the local fluid velocity while maintaining fixed mass in a quasi-Lagrangian manner. The gas cells exchange fluxes of energy, momentum, and magnetic flux with their nearest neighbors in a conservative, finite-volume Godunov-like fashion across the ``effective faces" defined by a kernel-weighted volume partition and a weighted least-squares gradient matrix (see \citealt{hopkins2015_gizmo} for full expressions). 

\subsection{Gravity}

The gravitational acceleration and tidal field are computed with {\small GIZMO}'s approximate Barnes-Hut oct-tree solver \citep{Springel_2005_gadget,hopkins2015_gizmo}, modified to enforce a maximum node opening angle $\Theta < 0.5$ in addition to the other tree opening criteria, to control the error in the external force on dense subsystems (gas clumps, clusters, binaries) whose internal self-gravity is much stronger than the external field \citep{grudic_2020_cluster_formation}. The gravity calculation for gas cells is optimized by the \citet{grudic:2021.adaptive} adaptive force-updating scheme, calling the gravity solver only as frequently as needed and using a predictor to estimate the field between calls (setting the update frequency parameter $q_{\rm f}$ defined in \citealt{grudic:2021.adaptive} to $0.0625$).
 
Gravitational softening for gas-gas interactions is fully adaptive, scaled to the local inter-cell spacing at all times, with additional terms to ensure conservation \citep{price_monaghan_softening}. The gravitational softening length (radius of the compact softening spline) for star-star interactions is fixed at $\rm 18 AU$, and the effective softening length used for gas-star interactions is taken to be the greater of the gas cell's or the star particle's softening length. The use of softening for stars makes the dynamics of binaries and encounters with periastron $<18 \rm AU$ unphysical.  

\subsection{Timestepping}

We advance the gas and stars in time using {\small GIZMO}'s adaptive, hierarchical powers-of-two individual block timestepping scheme \citep{Springel_2005_gadget}. To control the orbital integration accuracy we use the \citet{grudic:2020.tidal.timestep.criterion} tidal timestep criterion, taking the accuracy parameter to be $\eta=0.01$. Stars obey additional timestep criteria designed to anticipate stellar encounters and give good conservation in binary integration. Gas cells and stars also obey a set of additional timestep criteria designed to anticipate the arrival of feedback.

We integrate gas cells with to the usual 2nd-order kick-drift-kick integrator, while stars use a modified version of the 4th-order Hermite integrator \citep{makino:hermite} to achieve the level of accuracy necessary to handle close encounters and preserve binary orbits over the $\sim 10 \rm Myr$ duration of the simulation.

\subsection{Thermodynamics}
\label{sec:thermo}
We use a gas equation of state that accounts for the varying adiabatic index due to the varying ratio of para- to ortho-hydrogen \citep{vaidya_2015_eos}, and variations in the fraction of molecular H. The temperature and ionization state of the gas are evolved using a standard implicit method, operator-split with the MHD evolution, accounting for various cooling and heating processes. These processes include molecular and fine-structure cooling, cosmic ray heating, dust cooling and heating (coupled to the radiation field), photoelectric heating (assuming a fixed 1.7 Habing background \citep{draine_1978_isrf} attenuated with a 6-bin {\small TREECOL} column density estimator, \citealt{treecol}, plus local stellar irradiation from the RT solver), metal line cooling, H photoionization (coupled to the radiation field), and collisional ionization of H and He. The molecular fraction of H is evolved explicitly according to a simplified H-only network accounting for the local photodissociation rate due to cosmic rays and Lyman-Werner photons from the assumed background radiation field, and irradiation by stars in the simulation (see \S\ref{sec:radiation}). Detailed formulae and fitting functions for radiative cooling and heating processes are given in \citet{fire3}.

\subsection{Sink particles}
\label{sec:sinks}

Stars and protostars are represented by sink particles in the simulation, which are converted on-the-fly from gas cells that satisfy a number of checks designed to identify physical centres of collapse that will exceed the effective resolution limit of the simulation \citep[e.g.][]{Bate_1995_accretion,Krumholz_2004_sinks_in_eulerian, Federrath_2010_sinks,hopkins2013_sf_criterion,gong_2013_athena_sinks}. Sink particles can accrete nearby gas cells whose centres of mass lie within the accretion radius $R_{\rm sink}=18 \rm AU$ and satisfy various other checks. When a gas cell is accreted, the position, velocity, and internal angular momentum of the sink are updated to conserve centre of mass, total momentum, and total angular momentum to machine precision. Sink particles can merge only if they are bound to each other with a semimajor axis $<R_{\rm sink}$, and the lesser sink mass is $<10\Delta m$ where $\Delta m=10^{-3}\,\msun$ is our nominal mass resolution 
The vast majority of sinks never satisfy the merging criteria during the simulation, so the main results do not rely on the particulars of the merging strategy.


\subsection{Stellar evolution}
Each sink particle contains a star that accretes continuously from an internal mass reservoir fed by the resolved sink accretion process described in \S\ref{sec:sinks}. The luminosity, temperature, and radius of each star are each explicitly evolved in turn according to the sub-grid protostellar evolution prescription originally implemented in the {\small ORION} code by \citet{Offner_2009_radiative_sim}. This model integrates the protostellar evolution through a sequence of phases, ending on the main sequence. Note that massive stars formed in our simulations routinely ignite H while still accreting appreciably, and move along the main sequence thereafter. The zero-age main sequence (ZAMS) mass $M_{\rm ZAMS}$ used for determining the stellar lifetime, modeling feedback rates, and measuring the IMF is taken to be the {\it greatest} mass that the star ever has, after H burning has begun.  

\subsection{Feedback}
We account for stellar feedback in the form of accretion- and fusion-powered stellar radiation, stellar winds, protostellar jets, and core-collapse supernovae.
\subsubsection{Radiative transfer}
\label{sec:radiation}
The radiation field is evolved in 5 frequency bins (H ionizing, FUV, NUV, optical-NIR, and FIR) using {\small GIZMO}'s M1 solver \citep{levermore_1984_M1, hopkins_grudic_2019, fire_RT}, in which gas cells exchange fluxes of radiation across effective faces, i.e. the same mesh-free volume discretization as used for MHD solver (\S\ref{sec:mhd}). To make the calculation tractable, we assume a reduced speed of light $\tilde{c} = 30 \rm km\,s^{-1}$, sufficient to capture the dynamics of D-type HII region expansion \citep{geen_2015_rt,starforge.methods}.

Stars act as sources, injecting photons into the simulation domain in all 5 bands, according to the spectral energy distribution determined by the stellar evolution model. Dust may also radiate photons into the FIR band. We account for scattering and absorption by dust in all 5 bands, and the absorption of Lyman continuum photons by HI. Absorbed ionizing photons are assumed to be promptly re-radiated isotropically in the optical-NIR band, and radiation absorbed in all other bands is assumed to be re-radiated by dust in the FIR band. The radiation field couples to the fluid via dust heating/cooling, photoelectric and photoionization heating, and radiation pressure terms in the energy and momentum equations.

\subsubsection{Protostellar jets}

We model protostellar jets using the prescription of \citet{Cunningham_2011_outflow_sim}, wherein a fraction $f_{\rm w}=0.3$ of sink particle accreta is diverted into a jet, which is launched in a collimated pattern along the sink angular momentum axis with a speed $v_{\rm jet} = f_{\rm K} \sqrt{G M_\star/R_{\star}}$, with $f_{\rm K} = 0.3$. Note that these jet parameters have the greatest influence upon the IMF of any parameter choice in our simulation that we have investigated \citepalias{starforge_jets_imf}, and our adopted parameters are similar to those adopted in other studies that have used this model \citep{Cunningham_2011_outflow_sim,hansen_lowmass_sf_feedback,krumholz_2012_orion_sims, Offner_2016_jets,Cunningham_2018_feedback, murray_2018_jets,rosen_2020_jets_radiation}, and result in outflow masses and momenta that match observational constraints \citep{matzner_mckee_2000_jets, Cunningham_2011_outflow_sim, maud:2015.outflows}. The jets are injected as new gas cells with mass $\Delta m_{\rm w} = 10^{-4} M_\odot = 0.1 \Delta m$ spawned near the sink in pairs with opposite positions and velocities, conserving center of mass and momentum to machine precision. 

\subsubsection{Stellar winds}
Winds from $>2M_\odot$ main-sequence stars are modelled using the following prescription for the mass-loss rate\footnote{The corresponding equation in \citetalias{starforge.methods} contains an error in the numerical prefactors, corrected here.}:
\begin{equation}
   \frac{\dot{M}_{\rm wind}}{M_\odot\rm yr^{-1}} = \min \left(10^{-15} L_{\rm MS}^{1.5}, 10^{-22.2} L_{\rm MS}^{2.9} \right)  Z_{\rm \star}^{0.7},
   \label{eq:mdotwind}
\end{equation}
where $L_{\rm MS}$ is the ZAMS luminosity for a given stellar mass, from \citet{tout_1996_mass_lum}. This models the expected metallicity dependence of line-driven winds, the ``weak wind problem" for B and late O dwarfs, and a mass loss rate for early O stars that is roughly $\sim 3\times$ less than the widely-used \citet{vink_2001_winds} prescription. This conservative estimate of $\dot{M}_{\rm wind}$ is motivated by the observation of mass loss rates $\sim 2-3\times$ less than predicted by theory \citep{smith_2014_winds}. The terminal wind velocity varies with the escape speed and effective temperature following \citet{lamers_1995_wind_vesc}, modeling temperature-dependent bi-stability jumps. Stars with masses $>20 M_\odot$ can evolve to a Wolf-Rayet phase once reaching a mass- and metallicity-dependent age fit to \citet{meynet_2005_wolfrayet}, which we model by enhancing their mass-loss rates by a factor of 10 compared to Eq. \ref{eq:mdotwind}. Note that our assumptions about stellar evolution are not fully consistent with our assumptions about mass loss - a more realistic and self-consistent mass loss and evolution prescription is desirable, but beyond the present scope.

Winds are implemented numerically either by spawning new gas cells, or by injecting the appropriate mass, energy, and momentum into surrounding gas according to the conservative weighting scheme given in \citet{Hopkins_2018_sne_feedback}, depending on whether the free-expansion radius is resolvable.

\subsubsection{Supernovae}
Stars with $M_{\rm ZAMS} >8M_\odot$ in the simulation end their lives as a core-collapse supernova, with a mass-dependent lifetime given by \citetalias{starforge.methods} Eq. 34. The ejecta are injected directly into the simulation as resolved shells of gas cells at the 18AU sink radius with the fiducial $10^{-3}M_\odot$ mass resolution. The ejecta cells are then followed self-consistently through the free-expansion phase onward 
We assume the ejecta are isotropic and have a total kinetic energy of $10^{51} \rm erg$.

\subsection{Initial conditions and setup}
\begin{table}
    \centering
    \setlength\tabcolsep{2.0pt}
    \begin{tabular}{c|c|c|c}
        Symbol & Meaning & Expression & Init. Value  \\
        \hline
       
        $M_{\rm}$ & Cloud mass & -- & $2\times 10^4 M_\odot$ \\ 
        $R_{\rm}$ & Cloud radius & -- & $10 \rm pc$ \\
        $L$ & Box size & $10R$ & 100pc \\
        $n_{\rm H}$ & Number density of H nuclei & $3X_{\rm H} M /\left(4\uppi R^3 m_{\rm p}\right)$ & $146 \rm cm^{-3} $ \\ 
        $t_{\rm ff}$ & Free-fall time & $\uppi \sqrt{R^3 / (8 G M)}$ & 3.7Myr \\ 
        $\Sigma$ & Mean surface density & $M/\left(\uppi R^2\right)$ & $64 M_\odot \, \rm pc^{-2}$ \\
        $\sigma_{\rm 3D}$ & 3D velocity dispersion & -- & $2.9 \rm km\,s^{-1}$ \\
        $\alpha_{\rm turb}$ & Turbulent virial parameter & $5 \sigma_{\rm 3D}^2 R_{\rm }/\left(3 G M_{\rm }\right)$  &  2\\
        $\mathcal{M}$ & Turbulent Mach number & $\sigma_{\rm 3D}/c_{\rm s}$ & $15$ \\ 
        $T$ & Temperature & -- & $20 \rm K$ \\ 
        $e_{\rm rad}^{\rm FIR}$ & FIR energy density & -- & $0.3 \rm eV \, cm^{-3}$ \\ 
        $u\left(6-13.6\rm eV\right)$  & FUV energy density & -- & 1.7 Habing \\ 
        $B$ & Magnetic field strength  & -- & 2 $\mu \rm G$\\
        $\mu_{\rm 0}$ & Norm. mass-to-flux ratio & $0.4 \frac{G^{1/2} M}{\uppi R^2 B_{\rm 0}}$ & 4.2 \\
        $M_{\rm mol}$ & Mass of molecular gas & -- & 0 \\
    \end{tabular}
    \caption{Summary and glossary of parameters of the simulated cloud and their initial values (subscripted with $_0$ throughout this paper when referring to the respective initial values). Note that $\mathcal{M}$ assumes $c_{\rm s}=0.2 \rm km\,s^{-1}$, but $c_{\rm s}$ varies self-consistently according to the gas's thermodynamic evolution (\S\ref{sec:thermo}), and this value only represents the mean $\sim 10 \rm K$ temperature of dense gas.
    }
    \label{tab:ics}
\end{table}

The initial parameters of the simulation are summarized in Table \ref{tab:ics}. The simulation domain is a $L=100\rm pc$ periodic box. The GMC is initially a uniform-density sphere with mass $M_{\rm 0}=2\times 10^4 M_\odot$ and radius $R_{\rm 0}=10 \rm pc$, placed at the center of the box. These parameters were selected to match the typical mean surface density of GMCs in the Solar neighborhood \citep[e.g.][]{lada:2020.sigma.gmc.const}. The rest of the box is filled with gas with $1/1000$ the density of the cloud, containing a total gas mass of $\sim 5000 M_\odot$, and the gas mass throughout the domain is discretized into $\sim 25$ million $10^{-3} M_\odot$ gas cells. The cloud is given an initial pseudo-turbulent random velocity field constructed in Fourier space to have a $\propto k^{-2}$ power spectrum with a natural mixture of compressive and solenoidal modes (i.e. $E_{\rm sol}=2E_{\rm comp}$), normalized to give a virial parameter $\alpha_{\rm turb}=2$ \citep[e.g.][]{bate2003}. The diffuse medium is initially static. All gas is initially of Solar composition and all H is initially atomic.

The magnetic field $\mathbf{B}$ is initially uniform in the $+z$ direction with a strength $B=2 \rm \mu G$, giving the cloud a normalized mass-to-flux ratio $\mu_{\rm 0}=4.2$ (where $\mu_{\rm 0}=1$ would be the critical threshold for collapse of a static sphere, \citealt{Mouschovias_Spitzer_1976_magnetic_collapse}). The initial FIR radiation field has an energy density of $0.3 \rm eV\,cm^{-3}$ and a black-body SED with a temperature of $20 \rm K$, modeling the dust emission component of the interstellar radiation field in the Solar neighborhood. The gas temperature is also initialized to $20 \rm K$, but gas quickly reaches a new equilibrium temperature based on local conditions, so our results are insensitive to the initial temperature.

\section{Results}
\label{sec:results}
\begin{figure*}
    \centering
    \includegraphics[width=1\textwidth]{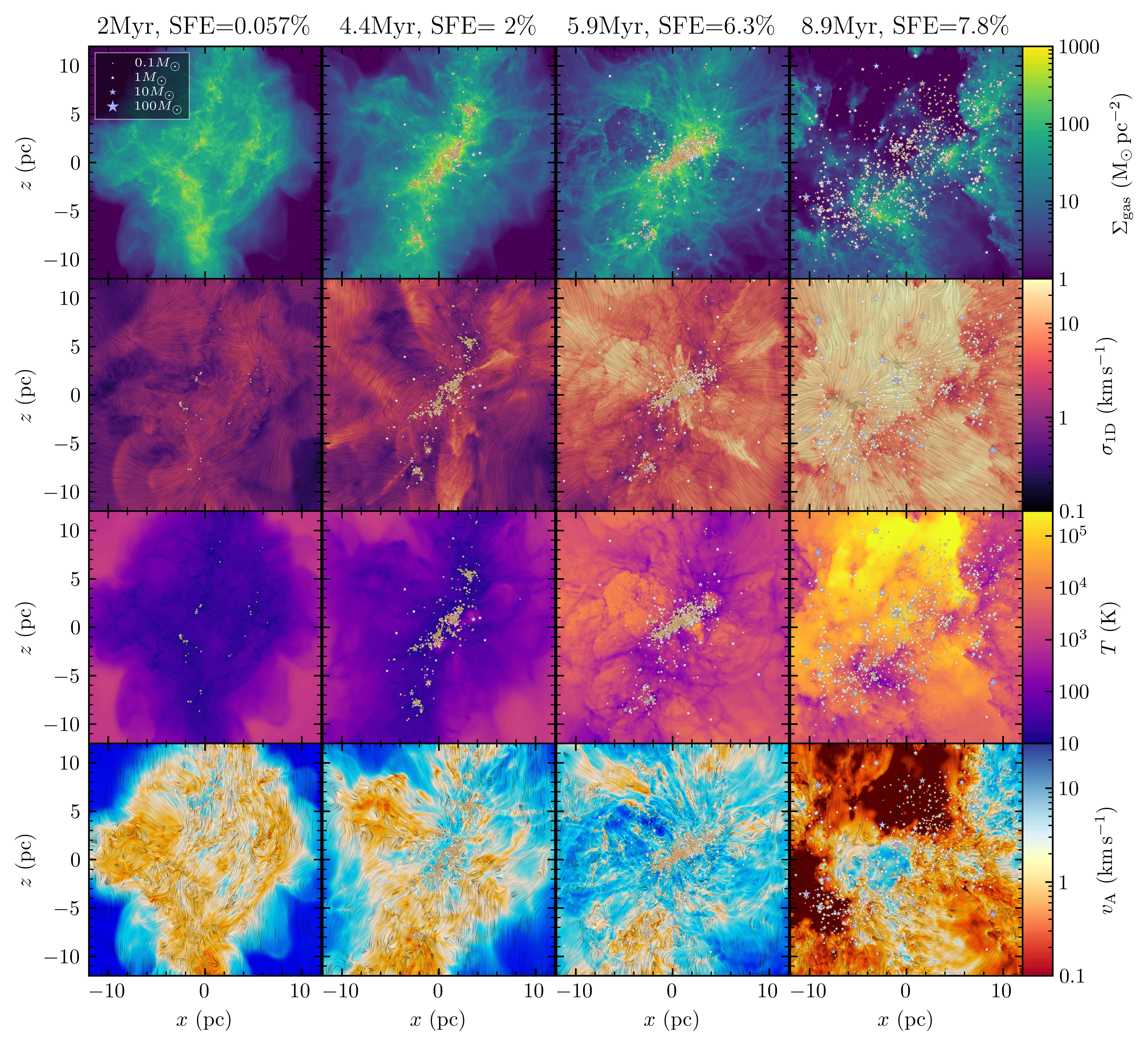}\vspace{-8mm}
    \caption{State of the cloud and star cluster at 4 different times (left to right) as visualized by the gas surface density $\Sigma_{\rm gas}$ (row 1), line-of-sight gas velocity dispersion $\sigma_{\rm 1D}$ (row 2), mass-weighted gas temperature $T$ (row 3) and the mass-weighted RMS Alfv\'{e}n speed $v_{\rm A}$ (row 4). Flowlines in row 2 plot the mean gas velocity perpendicular to the page, and flowlines in row 4 visualize the magnetic field as would be observed from the dust polarization angle. The positions of stars are indicated by markers whose size and colors correspond to their mass (see key in top left panel). A 3D animated rendering of the simulation is available \href{www.starforge.space/M2e4_fullphysics_flyaround.mp4}{here}.} 
    \label{fig:renders}
\end{figure*}
We ran the simulation on the Frontera supercomputer at the Texas Advanced Computing Center. It required 107 wall-clock days of runtime to run to 9.3Myr, for a total of 1.2 million core-hours. 
$\sim 160$ million timesteps were taken in total, but most elements in the simulation required significantly fewer cycles thanks to the code's adaptive block timestepping scheme. The shortest simulation timesteps were on the order of 1 day, generally for supernova ejecta, resolved Wolf-Rayet winds, and stars in hard massive binaries.

\subsection{Overview}

Figure \ref{fig:renders} visualizes the time evolution of the gas mass distribution, gas kinematics, gas temperature, and the magnetic field strength (via the RMS Alfv\'{e}n speed $v_{\rm A}=B/\sqrt{4 \uppi \rho}$) and dust polarization morphology, with the positions of stars superimposed. First the cloud and surrounding envelope quickly establish a thermal structure in equilibrium with the interstellar radiation field and cosmic ray background, with temperatures ranging from a few $10^3 \rm K$ in the warm ambient medium to $\sim 4 \rm K$ in the deepest parts of the cloud. From its initial uniform state, the random velocity field leads to shocks and internal density perturbations, which develop into a network of filaments and hubs. These dense regions go on to host the first gravitationally-unstable cores
, which collapse to form the first stars and subclusters (Figure \ref{fig:renders} column 1). 


Roughly 50\% of stars by number form by 4Myr (roughly one initial cloud free-fall time, Fig. 1 column 2). The rate of star formation increases as more subregions throughout the cloud contract enough to produce collapsing cores, and established protostars continue to accrete. The first massive stars have finished accreting by 4Myr, clearing out their environment via feedback and ionizing their immediate surroundings, but the influence of feedback on the morphology, kinematics, and thermal state of the cloud is still limited. The velocity map shows that the cloud is permeated by high-velocity outflows, but this difficult to see in the gas or dust mass-weighted morphology \citep{krumholz_2012_orion_sims,starforge_jets_imf}.

By 6Myr (Fig. 1 column 3), most of the eventual stellar mass has been accreted, star formation has slowed down, and the cloud is no longer gravitationally bound $(\alpha_{\rm turb} > 2)$. The morphology of the cloud is considerably disturbed by feedback-driven cavities of warm, photoionized gas, some of which have broken through the edge of the cloud to form champagne flows. At this time most subclusters have assembled into a single dense, primary, central cluster. 

The cloud continues to expand under the influence of feedback, opening a large central cavity through which warm gas and radiation escape. The star formation rate continues to drop in turn, and the total stellar mass accreted levels off at $\sim 1600 M_\odot$, for an integrated star formation efficiency of $8\%$. At 8.3Myr the first supernova occurs from a $31 M_\odot$ ZAMS progenitor. \footnote{A $31 M_\odot$ ZAMS star of solar composition would actually fail to form a supernova according to many current stellar evolution/explosion models, but here we adopt a simplified prescription wherein all $>8 M_\odot$ stars produce a core-collapse SN.}
The final column of Figure \ref{fig:renders} shows the immediate aftermath: the cloud morphology is highly disturbed, forming cometary structures from the interaction of the blast with the remaining dense clumps. The cluster has also expanded considerably from its former dense state. By the end of the simulation at 9.3Myr some star formation is still ongoing but at $<1\%$ of the peak rate. 

\subsection{Gas evolution}

\begin{figure}
    \centering
    \includegraphics[width=\columnwidth]{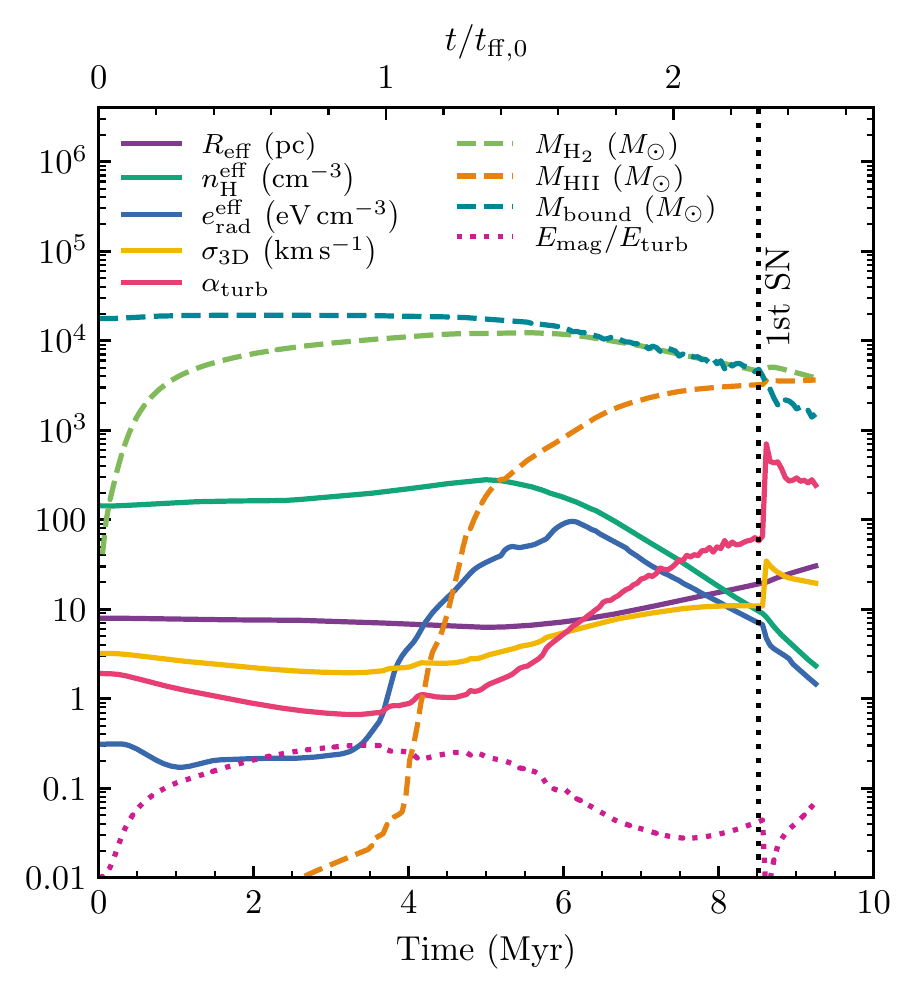}\vspace{-8mm}
    \caption{Evolution of different global properties of the gas distribution in the simulation: the half-mass radius $R_{\rm eff}$, half-mass average density $n_{\rm H}^{\rm eff}$ and radiation density $e_{\rm rad}^{\rm eff}$, velocity dispersion $\sigma_{\rm 3D}$, virial parameter $\alpha_{\rm turb}$, molecular gas mass $M_{\rm mol}$, ionized gas mass $M_{\rm HII}$, bound gas mass $M_{\rm bound}$, and the ratio of magnetic to turbulent energy.}
    \label{fig:global.gas.props}
\end{figure}

Figure \ref{fig:global.gas.props} plots the time evolution of various global gas properties, computed from only the subset of gas cells that were originally in the cloud and are not injected jet or wind material. The cloud half-mass radius $R_{\rm eff}$ remains roughly constant at $8 \rm pc$ throughout most of the cloud evolution: although much of the initial turbulence does decay at first, the virial parameter $\alpha_{\rm turb}=-2E_{\rm kin}/E_{\rm grav}$ (also plotted) is never significantly less than unity before feedback starts to drive it back up, preventing significant global contraction. Consequently, the half-mass volume-averaged density of H nuclei $n_{\rm H}^{\rm eff}$ stays in the range $150-200 \rm cm^{-3}$ until the cloud expands and the mean density drops.\footnote{The volume-averaged density should not be confused with the mass-weighted mean density here, which is generally considerably higher in this simulation ($\sim 10^4 \rm cm^{-3}$) due to the $\propto \mathcal{M}^2$ clumping factor of gas in compressible turbulence \citep[][]{vazquez_lognormal}.} Once feedback does become active, the cloud expands, accelerating to a velocity dispersion of $\sigma_{\rm 3D} \sim 10 \rm km\,s^{-1}$. The total magnetic energy within the cloud is initialized to only 1\% that of turbulence, but the magnetic field is rapidly amplified by the initial turbulent motions until the magnetic energy is $\sim 20\%$ of the turbulent energy. This results in a mass-weighted median field strength of $\sim 10 \mu \rm G$, comparable to Zeeman measurements in the Milky Way in the $100-10^3 \rm cm^{-3}$ density range that most of the gas in the simulation occupies \citep{crutcher:2012.magnetic.fields}.

The radiation field, measured as the volume-averaged radiation energy density within $R_{\rm eff}$, $e_{\rm rad}^{\rm eff}$, initially remains close to the background density of $0.3 \rm eV\,cm^{-3}$, because the luminosity of gas dissipation ($\sim 5 L_{\odot}$) and stellar accretion is small compared to the $\sim 3000 L_{\odot}$ required to sustain a comparable energy density. Eventually around 3Myr the total luminosity does cross this threshold and $e_{\rm rad}^{\rm eff}$ begins to rise, reaching a peak value of $100 \rm eV \,cm^{-3}$ at 6Myr when star formation is most intense and the star cluster is densest. It then decays roughly exponentially as star formation is quenched, the cluster disperses, and the cloud becomes more optically thin. It is worth noting that, like the gas density, the radiation experienced by an average H nucleus or protostellar system can be significantly greater than this volume-averaged value \citep{lee.hopkins:2020.sf.radfield}.

Lastly, Figure \ref{fig:global.gas.props} plots the evolution of various mass components of the cloud. The cloud is initially entirely bound by self-gravity, but the total bound mass $M_{\rm bound}$ (defined as gas with negative total energy in the rest frame) begins to drop after the onset of feedback, reaching $\sim 10^3 M_\odot$ by the end of the simulation. The cloud is initially atomic, but turns mostly molecular within the first 3Myr, reaching a peak molecular mass $M_{\rm mol}=1.2\times10^4 M_\odot$, 60\% of the total mass. The molecular mass declines after the peak of star formation at 6 Myr, when the cloud is dispersed and becomes increasingly photodissociated and photoionized by starlight and the interstellar radiation field. However, a significant ($4000 M_\odot$) amount of molecular mass is still present at the end of the simulation, mainly in the surviving self-shielding dense clumps that often continue forming stars. The ionized mass $M_{\rm HII}$ increases rapidly around 4Myr as the first $>20 M_\odot$ stars with significant ionizing luminosity form,  eventually ionizing $20 \%$ of the cloud mass.

\subsection{Star cluster evolution and kinematics}
\begin{figure}
    \centering
    \includegraphics[width=\columnwidth]{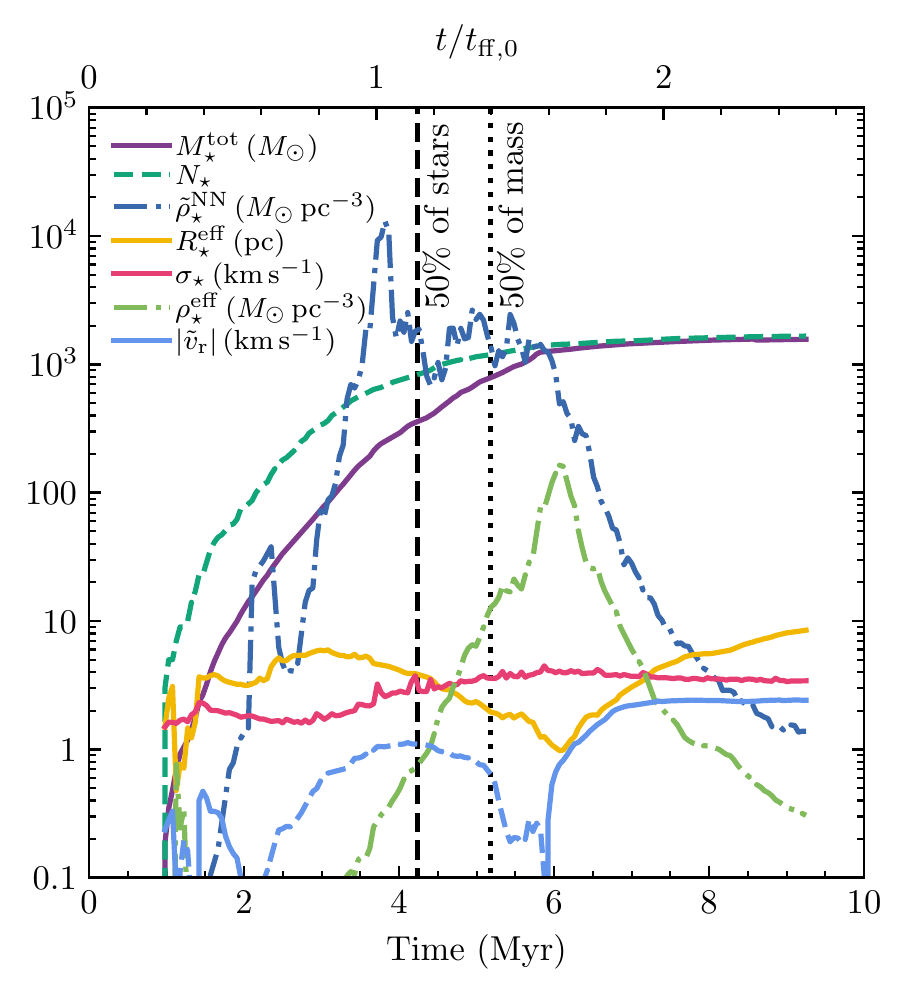}\vspace{-8mm}
    \caption{Evolution of different global properties of the star cluster in the simulation: total stellar mass $M_{\rm \star}^{\rm tot}$, number of stars $N_{\rm star}$, median nearest-neighbor-estimated stellar density $\rho_{\star}^{\rm NN}$, half-mass radius $R_{\rm star}^{\rm eff}$, velocity dispersion $\sigma_{\rm \star}$ (neglecting binary motion), half-mass average stellar density $\rho_{\star}^{\rm eff}$, and median radial velocity with respect to the median stellar velocity and position, $\tilde{v}_{\rm r}$. $\tilde{v}_{\rm r}$ is plotted with a dashed curve when negative and a solid curve when positive. Vertical lines indicate the times at which $50\%$ of the cluster has formed by number and by mass respectively.}
    \label{fig:global.star.props}
\end{figure}
Figure \ref{fig:global.star.props} plots various global properties of the star cluster as a function of time. The cluster grows in mass and number of stars until star formation is quenched, and the number of stars rises significantly sooner than the total mass in stars -- there is a characteristic time lag of $\sim 1 \rm Myr$ between the number-weighted and mass-accretion-weighted median star formation time (shown as vertical lines on Fig. \ref{fig:global.star.props}), because many stars require a non-negligible amount of time to acquire their mass once formed (see \ref{sec:tform} for a detailed analysis).

Unlike the gas, the star cluster undergoes significant collapse, contracting in size by factor of $\sim 5$ to a minimum half-mass radius $R_{\rm \star}^{\rm eff}\sim 1 \rm pc$ at 6Myr. 
This implies that the spatial and kinematic stellar properties do {\it not} simply trace that of the gas. Rather, the stars form preferentially in the dense, infalling regions of the cloud, which are necessarily regions that has predominantly compressive motions. This apparently imprints upon the cluster kinematics in turn.




The primary cluster assembles from a collection of subclusters, in a hierarchical fashion \citep{bonnell:2003.hierarchical,grudic_2017}. Like \citet{bonnell:2003.hierarchical} we compute two different stellar density statistics, 1. the stellar half-mass volume-averaged density $\rho_{\rm \star}^{\rm eff}$, and 2. a median local stellar density $\rho_{\star}^{NN}$, the median volume-averaged density of stars within a sphere enclosing the 10 nearest neighbors of each star. $\tilde{\rho}_{\star}^{\rm NN}$ is many orders of magnitude greater than $\rho_{\rm \star}^{\rm eff}$ during the initial contraction and hierarchical assembly of the cluster -- in fact the relative evolution of the two densities looks almost identical to that reported in \citet{bonnell:2003.hierarchical}, but rescaled to our different GMC model, which has lower density and a longer dynamical time. This suggests that the various additional physics we consider here do not seriously alter this picture of cluster assembly, at least up to the point where feedback is important. 


The assembly of the cluster coincides with the expulsion of gas from the central region by feedback, so the cluster never has a chance to virialize into a structure that is globally bound by stellar self-gravity. Rather, stars on unbound trajectories reach periapsis and then continue outward, so the cluster re-expands with a radial velocity on the order of the stellar velocity dispersion $\sigma_{\rm \star}$. 

\begin{figure}
    \centering
    \includegraphics[width=\columnwidth]{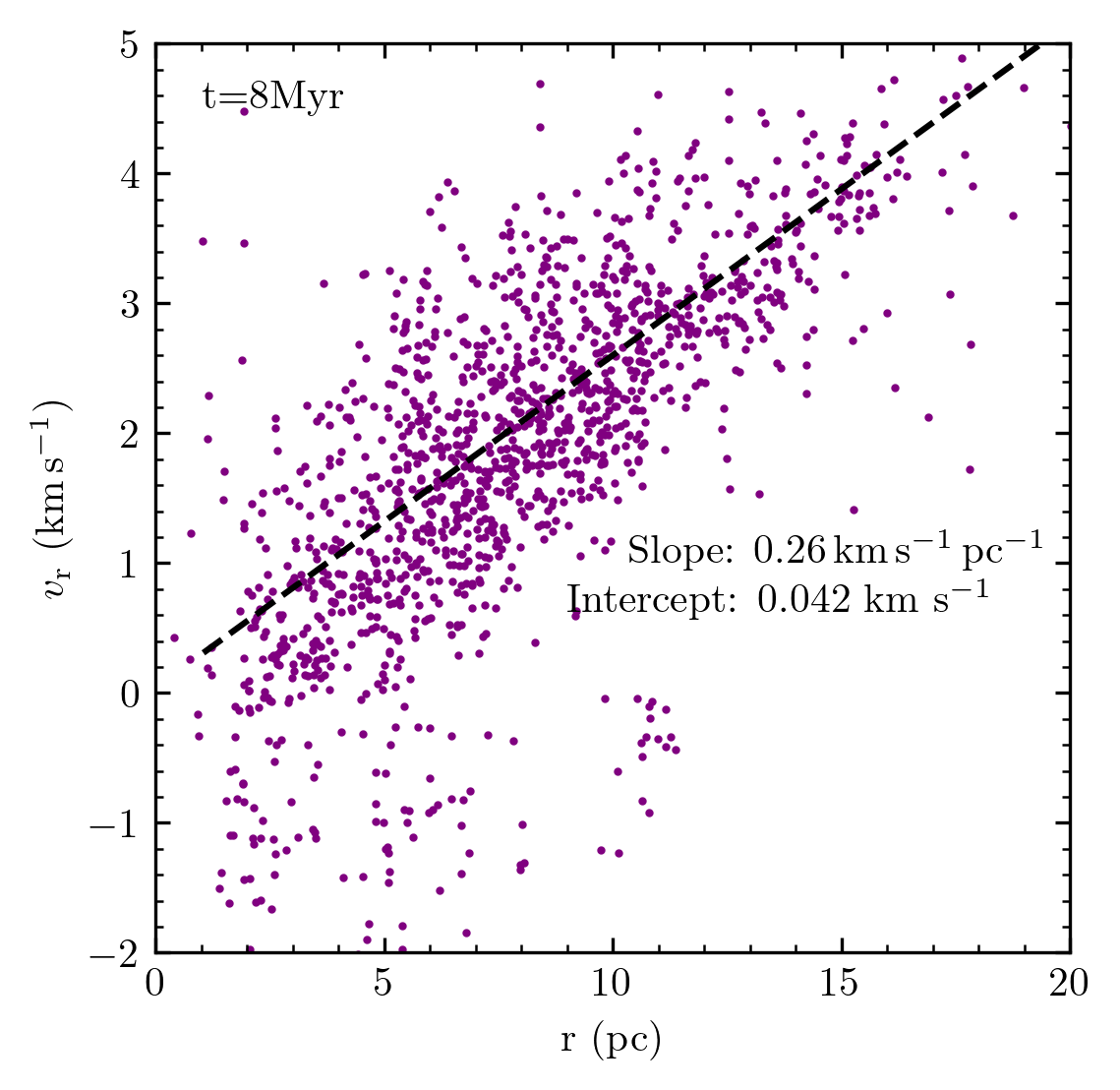}\vspace{-8mm}
    \caption{Radial velocity of stellar systems as a function of distance $r$ from the median stellar position at 8Myr in the simulation, measured from the mass-weighted median stellar position. We perform a robust linear fit to fit the relation of the primary cluster (dashed), with parameters given. The primary cluster exhibits ``Hubble-like" kinematics ($v_{\rm r} \propto r$) at this time.}
    \label{fig:r_vs_vr}
\end{figure}

Figure \ref{fig:r_vs_vr} shows that, after some time, the free expansion of the cluster from this dense configuration assumes a ``Hubble-like" relation between radius and radial velocity  $v_{\rm r} \propto r$, as seen in some expanding young star clusters in our Galaxy \citep{kuhn:2019.gaia.cluster.kinematics}. The interpretation of this relation within the context of the simulation is straightforward: the radius of an unbound star originating in the central cluster evolves as $r  \approx v_{\rm r} t$, where $v_{\rm r}$ is its original radial velocity. The overall median outward radial velocity is $2 \rm km \,s^{-1}$, within the range measured by \citet{kuhn:2019.gaia.cluster.kinematics}, and the fitted slope of the $r-v_{\rm r}$ relation is $\sim 0.3\rm km\,s^{-1}\,pc^{-1}$. \citet{kuhn:2019.gaia.cluster.kinematics} estimated that at least 75\% of the star clusters in their sample were expanding, broadly consistent with estimates that bound cluster formation accounts for only 4-14\% of star formation in the Solar neighborhood \citep{goddard:2010.cfe}. Hence the fate of the cluster in the simulation may be typical for Solar neighborhood conditions. 


A more detailed analysis of the virial state, merger history, and influence of gas evacuation due to feedback, for this simulation as well as others, is presented in \citep{guszejnov.starforge.clusters}.

\subsection{Feedback rates}

\begin{figure}
    \centering
    \includegraphics[width=\columnwidth]{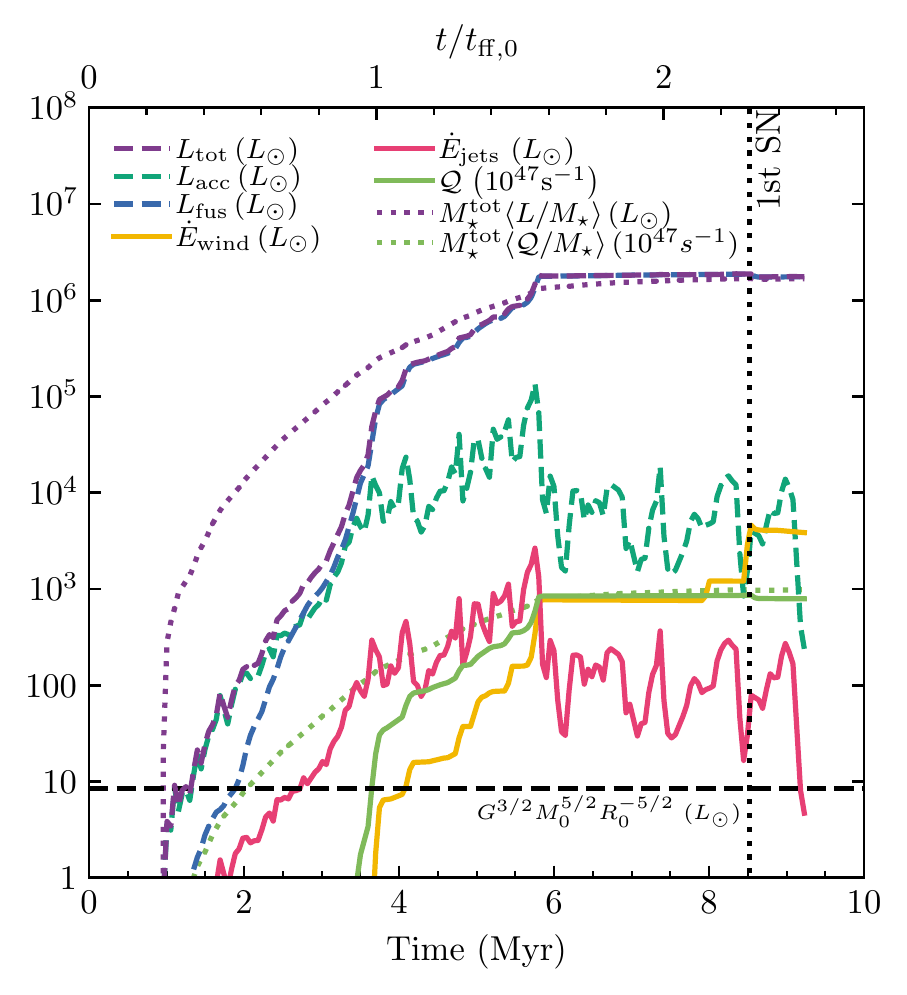}\vspace{-8mm}
    \caption{Evolution of the various energy injection rates from different feedback components: the total, accretion-powered, and fusion-powered radiative luminosities $L_{\rm tot}$, $L_{\rm acc}$ and $L_{\rm fus}$, the mechanical luminosity of winds $\dot{E}_{\rm wind}$ and jets $\dot{E}_{\rm jets}$, and the production rate of H ionizing photons $\mathcal{Q}$. For comparison we also plot the mean bolometric luminosity and ionizing photon production rate expected from a well-sampled IMF for a cluster of equal mass (dash-dotted), the time of the first supernova (dotted), and the characteristic luminosity of the cloud (dashed). }
    \label{fig:fb_energy}
\end{figure}
\begin{figure}
    \centering
    \includegraphics[width=\columnwidth]{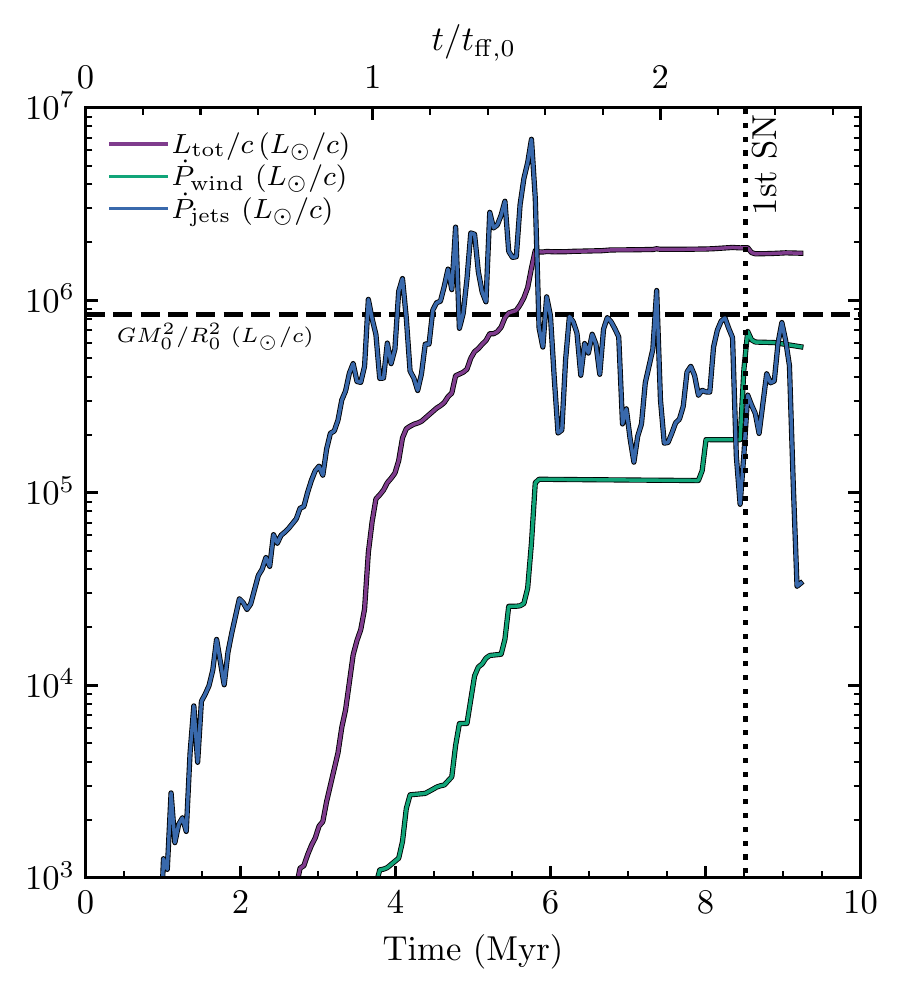}\vspace{-8mm}
    \caption{Evolution of various momentum injection rates (forces) from different feedback components: the single-scattered radiation force $L_{\rm tot}/c$, and the momentum injection rates of winds $\dot{P}_{\rm wind}$ and jets $\dot{P}_{\rm jets}$. For comparison we also plot the characteristic weight of the cloud (dashed) and the time of the first supernova (dotted).}
    \label{fig:fb_momentum}
\end{figure}

Figures \ref{fig:fb_energy} and \ref{fig:fb_momentum} plot the evolution of various stellar feedback input rates from the star cluster. These represent the ``raw" feedback rates injected from the stars, taking the energy injection rates of winds and jets to be the respective mechanical luminosities $\dot{E}=\dot{M} v^2/2$ and the momentum injection rates to be $\dot{P}=\dot{M} v$. Although we plot these quantities on the same axes to compare their evolution, different feedback mechanisms couple in different ways, so we caution that their relative importance can only be discerned at an order-of-magnitude level in such a diagram. The ionizing photon rate $\mathcal{Q}$ should not be quantitatively compared with other curves; we include it in the plot to indicate relative changes in the production of ionizing radiation and to compare with that expected from a well-sampled IMF.

Figure \ref{fig:fb_energy} shows the accretion power $L_{\rm acc} = \sum 0.5 G \dot{M} M_{\rm \star} / R_{\rm \star}$ is the dominant source of radiation for the first 2 Myr after the beginning of star formation, but once massive stars form the total luminosity is dominated by fusion (H, D, He) power, $L_{\rm fus}$,  which is calculated according to our stellar evolution model. The characteristic luminosity of the cloud is $L_{\rm 0} \sim G^{3/2}M_{\rm 0}^{5/2} R_{\rm 0}^{-5/2} \sim 10 L_\odot$, and all radiative and mechanical luminosities exceed this well before the cloud shows evidence of disruption by feedback. This implies that cooling is efficient for all feedback mechanisms, and hence feedback is best characterized by the {\it momentum} it imparts \citep{fall2010}, consistent with the findings of previous feedback simulations \citep{grudic_2016, rosen_2020_jets_radiation,lancaster:2021.wind.sims}. 

Figure \ref{fig:fb_momentum} shows that, among the different momentum injection rates, $\dot{P}_{\rm jets}$ is greatest during most of the star formation history, exceeded by photon momentum only at $\sim 6 \rm Myr$ when the most massive stars have finished forming and the star formation rate drops rapidly. It also has the greatest peak momentum output of all feedback channels, briefly reaching a rate nearly $10 \times$ the characteristic weight of the cloud $\sim GM_{\rm 0}^2/R_{\rm 0}^2$. However, in \citetalias{starforge_jets_imf} we found that jet feedback alone could  not fully disrupt the cloud and quench star formation in this GMC model (although it could in less-massive clouds). So although jets do play an important role in regulating star formation \citep[see also][]{nakamura_jets,wang_2010_jets,hansen_lowmass_sf_feedback,federrath_2015_inefficient_sf, Cunningham_2018_feedback}, jet feedback is not responsible for disrupting the cloud here. Such inefficient coupling of the available momentum may be due to internal momentum cancellation within the cloud and/or mismatch between the effective coupling scale of jet feedback and the cloud scale, or jet material escaping through cavities. It may also be that jet feedback has a more self-regulating nature less likely to overshoot the amount of feedback needed to disrupt the cloud, because it is proportional to the star formation rate, which responds directly to the dynamical state of the cloud, whereas radiation and winds do not. 

The cloud disruption is therefore due to some combination of radiation pressure, pressure of photoionized gas, and stellar winds. The peak momentum injection rate from photons $L_{\rm tot}/c$ is on the order of the weight of the cloud, so radiation pressure alone can conceivably disrupt the cloud. Note that the flux of H ionizing photons ($\mathcal{Q}\sim 10^{50}\,s^{-1}$) also ionizes a significant fraction of the total gas mass, whose expansion may also contribute significantly to cloud disruption. The wind momentum injection rate is also eventually comparable to the cloud weight, but only once the most massive stars enter their Wolf-Rayet phase $\sim 2 \rm Myr$ after the cloud has already started to expand. To disentangle the respective roles of massive stellar feedback in cloud disruption more conclusively, we must analyze our extended simulation suite, disabling individual mechanisms in turn (Guszejnov et al., in prep.).


Lastly, we estimate the radial momentum imparted by the supernova: Figure \ref{fig:global.gas.props} shows that it boosts the velocity dispersion $\sigma_{\rm 3D}$ (by then dominated by radial motion) from $\sim 10$ to $\sim 20\rm km\,s^{-1}$, for a total momentum $P \sim M_{\rm 0} \left(\Delta \sigma_{\rm 3D}\right) \sim 2\times 10^5 M_\odot\,\rm km\,s^{-1}$. This is close to the $P \sim 5\times 10^5 n_{\rm H}^{-1/7} M_\odot \,\rm km\,s^{-1}$ predicted by previous single supernova remnant simulations with more idealized setups (e.g. \citealt{cioffi_1988_snr, martizzi_2015_sne, gentry_2016_snr, Hopkins_2018_sne_feedback} and additional references therein), despite the more complex geometry.

\subsection{Star formation efficiency}
\label{sec:sfe}

\begin{figure}
    \centering
    \includegraphics[width=\columnwidth]{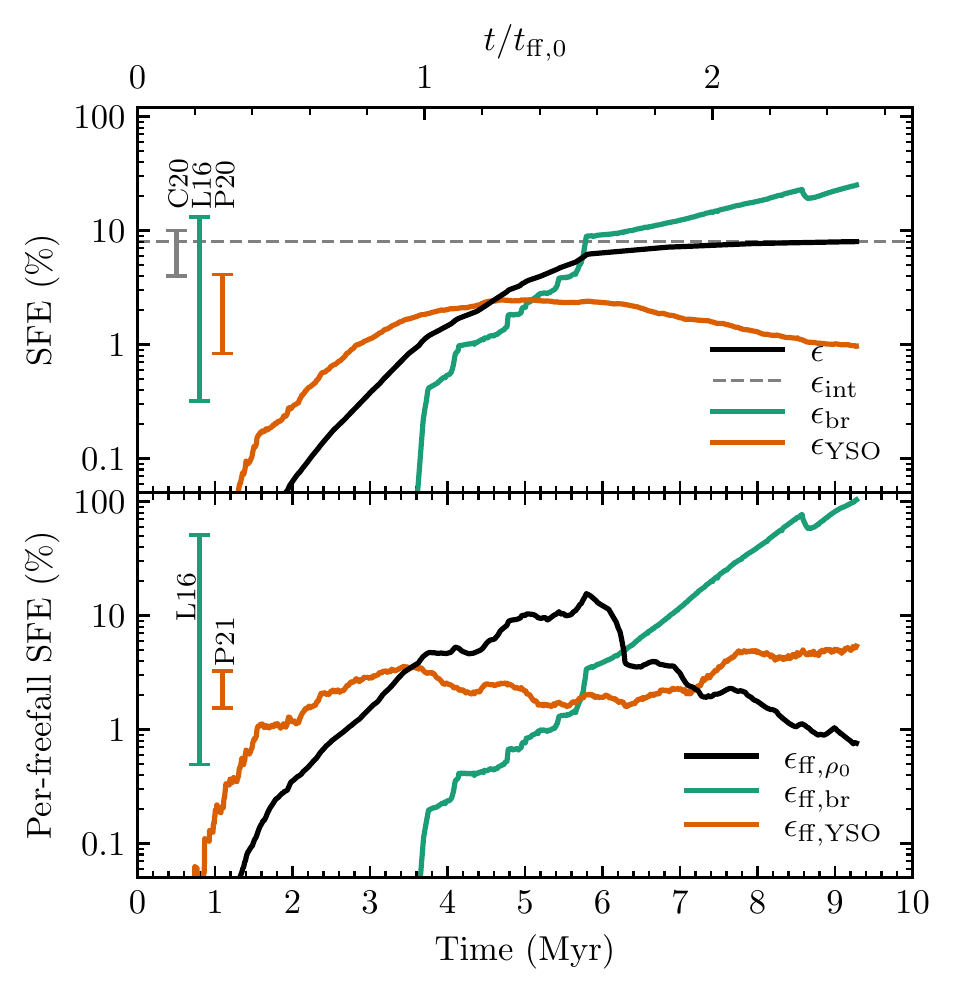}\vspace{-8mm}
    \caption{Evolution of various forms of star formation efficiency in the simulations and their observational proxies, including integrated (top) and per-freefall (bottom) flavors. We plot the true SFE $\epsilon=M_{\rm \star}^{\rm tot}\left(t\right)/M_{\rm 0}$, the final integrated SFE $\epsilon_{\rm int}$, the per-freefall SFE $\epsilon_{\rm ff,\rho_{\rm 0}}$, and their observational proxies from free-free emission and YSO-counting, defined in \S\ref{sec:sfe}. Error bars plot the $\pm \sigma$ ranges of observed values from different SFE studies: \citet{chevance_2020_gmcs} (C20), \citet{eve_lee_2016_GMC_sfe} (L16), and \citet{pokhrel:2020.gmc.sfe,pokhrel:2021.gmc.sfe} (P20,P21), color-coding the respective comparable simulated and observed quantities.} 
    \label{fig:SFE} 
\end{figure}

We now examine various measures of star formation efficiency that can be defined in the simulation. Observational estimates of star formation efficiency have considerable uncertainty, but here we are able to assess the accuracy of these definitions compared with the true efficiency.

Most basically one may ask what fraction of the initial gas mass $M_{\rm 0}$ has been converted to stars at time $t$:
\begin{equation}
    \epsilon\left(t\right) = \int_0^t \dot{M}^{\rm tot}_{\star}\left(t'\right) \mathrm{d} t' / M_{\rm 0},
\end{equation}
where $\dot{M}_{\rm \star}^{\rm tot}\left(t'\right)$ is the total stellar accretion rate at time $t'$. Once star formation begins, $\epsilon$ rises rapidly at first (increasing tenfold from 2-3.5Myr), then rises less steeply, and finally levels off to a final value of $ \epsilon_{\rm int} = 8\%$. This is within the $1\sigma $ range of values inferred from statistical modeling of gas and SFR maps in nearby galaxies \citep{chevance_2020_gmcs}.

We also measure the {\it  per-freefall} star formation efficiency \citep{km2005}:
\begin{equation}
    \epsilon_{\rm ff}\left(t\right) = \frac{\dot{M}^{\rm tot}_{\star}\left(t\right)}{M_{\rm gas}\left(t\right)/t_{\rm ff,0}},
\end{equation}
where $t_{\rm ff,0} = \uppi/2 \sqrt{R_{\rm 0}^3/(2 G M_{\rm 0})} = 3.7 \rm Myr$ is the freefall time at the initial mean density of the cloud. The bottom panel of Figure \ref{fig:SFE} shows that this quantity varies considerably throughout in the simulation, increasing steeply early on, peaking at 18\%, dropping off steeply as the cloud starts to be disrupted at 6Myr, and then decaying more gradually thereafter with a $1/e$-folding time of $3 \rm Myr$.

Although $\epsilon$, $\epsilon_{\rm int}$, and $\epsilon_{\rm ff}$ are of theoretical interest, they are not directly observable, and the available observable SFE quantities have a complex relationship with the true values of interest, depending heavily upon the stage of cloud evolution \citep{feldmann.gnedin.dynamic.sfe, eve_lee_2016_GMC_sfe, geen_2017_sfe,koepferl:2017.synthetic.obs, grudic_2018_mwg_gmc}. Motivated by these works, we make some basic estimates of observables before comparing with SFE measurements in the literature. Note that these are not true synthetic observations, which require significantly more post-processing \citep[e.g.,][]{haworth_synthetic_obs_review} but should still capture the basic behaviours of the observables. 

We define $\epsilon_{\rm YSO}$ as the ratio of stellar mass traced by $<2 \rm Myr$ old young stellar objects (YSO) assuming an average  YSO mass of $0.5 M_\odot$ and given a total gas mass $M_{\rm gas}$:
\begin{equation}
    \epsilon_{\rm YSO} = 0.5 M_\odot \frac{ N_{\rm YSO}\left(<2\rm Myr\right)}{M_{\rm gas}}.
    \label{eq:epsyso}
\end{equation}
Since it is not straightforward to measure protostellar masses directly, observational studies commonly adopt an assumed YSO average mass \citep[e.g.,][]{evans_2009}, which is close to the mean of various proposed forms of the IMF \citep[e.g.][]{kroupa_imf,chabrier_imf}. We also define a per-freefall SFE from YSO counts: 
\begin{equation}
    \epsilon_{\rm ff,YSO} = 0.5 M_\odot \frac{N_{\rm YSO}\left(<0.5 \rm Myr\right) / 0.5 \rm Myr }{M_{\rm gas} / t_{\rm ff}^{\rm eff}},
    \label{eq:effyso}
\end{equation}
where the effective freefall time $t_{\rm ff}^{\rm eff} = \sqrt{3 \uppi / 32 G \rho_{\rm eff}}$ is computed from the time-varying half-mass volume-averaged density $\rho_{\rm eff}$. This can be compared to recent measurements from \citet{pokhrel:2021.gmc.sfe}. The age cuts of $<0.5 \rm Myr$ and $<2 \rm Myr$ in Eqs \ref{eq:epsyso}-\ref{eq:effyso} correspond to the commonly-assumed lifetimes of Class 0+I and II YSOs, respectively \citep{Dunhamppvii2014}. 

We also model SFE measurements based on the ratio of free-free emission tracing massive stars \citep{murray_rahman_ob_assoc} to CO emission tracing molecular gas. Analogous to \citet{eve_lee_2016_GMC_sfe}, we define
\begin{equation}
    \epsilon_{\rm br} = \frac{1.37\mathcal{Q} \langle m_{\ast} / q \rangle}{1.37\mathcal{Q} \langle m_{\ast} / q \rangle + M_{\rm mol}},
\end{equation}
and 
\begin{equation}
    \epsilon_{\rm ff,br} = \epsilon_{\rm br}\frac{t_{\rm ff}^{\rm eff}}{t_{\rm ms,q}},
\end{equation}
where $\mathcal{Q}$ is the rate of ionizing photon emission from the cluster (plotted in Fig. \ref{fig:fb_energy}), $\langle m_{\ast} / q \rangle = 1.6 \times 10^{-47} M_\odot \rm s^{-1}$ is the IMF-averaged ratio of stellar mass to ionizing flux for a ZAMS stellar population, $M_{\rm mol}$ is the molecular gas mass (used as a proxy for the CO-traced mass in \citealt{eve_lee_2016_GMC_sfe}), and $t_{\rm ms,q}=3.9\rm Myr$ is the ionizing flux-weighted mean stellar lifetime.  We caution that full chemical modelling is required to model the complex relationship between CO emission and molecular gas mass \citep[e.g.,][]{glover:2011.molecules.not.needed.for.sf, Offner_2013_CO, keating:co.h2.conversion.mw.sims}. For comparison with this simulation we consider systems from \citet{eve_lee_2016_GMC_sfe} with total mass $<10^5 M_\odot$.

Comparing these modelled SFE quantities in Figure \ref{fig:SFE}, we see that both YSO number-weighted and ionization-weighted SFE estimators have significant biases with respect to the true values. $\epsilon_{\rm YSO}$ overestimates $\epsilon$ at early times because the mean stellar mass is less than the assumed $<0.5M_\odot$ and underestimates it at later times because the first stars cease to be counted in the YSO sample. 
$\epsilon_{\rm br}$ has the opposite problem: it underestimates $\epsilon$ at early times because the massive stars that dominate the contribution to $\mathcal{Q}$ have not yet formed, and overestimates at late times because the molecular gas mass drops on a timescale shorter than the lifetimes of the massive stars, as the cloud is dispersed. $\epsilon_{\rm ff,br}$ has a similar bias with respect to $\epsilon_{\rm ff}$, but its divergence toward large values is exaggerated even further because $t_{\rm ff}^{\rm eff}$ is also increasing as the cloud becomes less dense.

Most interesting is the behaviour of $\epsilon_{\rm ff,YSO}$ with respect to $\epsilon_{\rm ff}$. It overestimates $\epsilon_{\rm ff}$ at early times due to the assumed average stellar mass, then underestimates it at intermediate times (3-7 Myr), and settles to a nearly-constant overestimate at late times as $t_{\rm ff}^{\rm eff}$ increases and $N_{\rm YSO}\left(<0.5 \rm Myr\right)$ decreases. The net effect is that, after 2Myr, {\it the various factors in $\epsilon_{\rm YSO}$ (Eq. \ref{eq:effyso}) conspire to compress $\epsilon_{\rm YSO}$ to a much narrower range than the physical per-freefall SFE}. Quantitatively, the variation in $\epsilon_{\rm ff,\rho_{\rm 0}}$ from 2-8Myr is 0.4dex, while the variation of $\epsilon_{\rm ff, YSO}$ is 0.13dex, a factor of 3 smaller. 
Thus, while \citet{pokhrel:2021.gmc.sfe} proposed that their $\epsilon_{\rm ff,YSO}$ measurements had less dispersion than other works due to observational errors inherent in using diffuse tracers of massive stellar emission (a conclusion supported by our analysis of $\epsilon_{\rm ff,br}$), our simulation shows that this technique could potentially be {\it underestimating} the scatter. In general our results show how YSO-based measurements may underestimate the true scatter in the per-freefall SFE by a factor of $\sim 3$.

Overall, our SFE analysis shows that if we model the manner in which SFE is measured in observations, the cloud traces a range of SFE values that is consistent with similar systems in the Milky Way -- an important test for the model. However, in future work the model should be tested more sensitively, by performing mock-observations of SFE quantities that use the actual pipelines to catalogue YSOs and map the gas. \citet{pokhrel:2021.gmc.sfe} and \citet{hu:2021.gmc.sfe} also performed a more-detailed $\Sigma_{\rm gas}$-dependent analysis of $\epsilon_{\rm ff}$, which would likely produce more detailed constraints on SFE on different scales.



\subsection{Stellar initial mass function}
\begin{figure}
    \centering
    \includegraphics[width=\columnwidth]{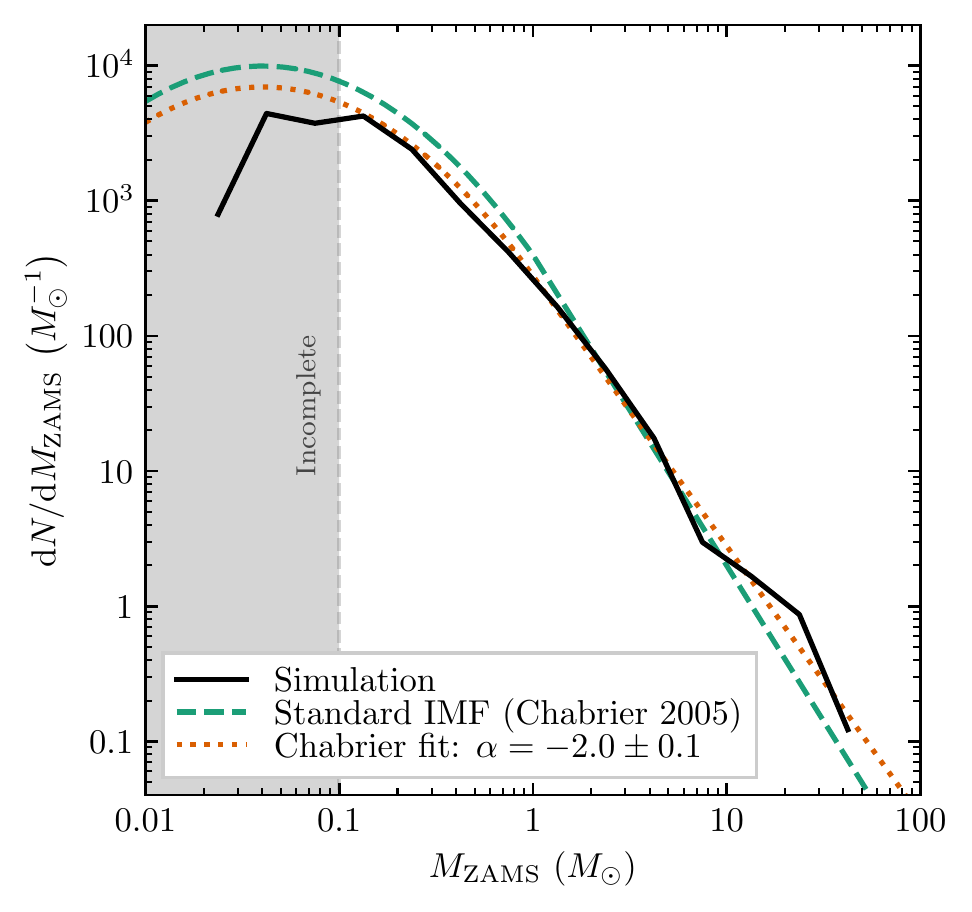}\vspace{-8mm}
    \caption{Stellar initial mass function ($\mathrm{d}N/\mathrm{d} M_{\rm ZAMS}$) predicted by the simulation. The shaded region indicates the mass range in which we expect low-mass incompleteness due to finite resolution \citep{bate_1997_resolution,starforge.methods}. We compare with the empirically derived IMF from \citet{chabrier_imf}, for the standard slope $\alpha=-2.35$, and to the maximum-likelihood fit $\alpha=-2.0\pm 0.1$ assuming stellar masses are independently sampled.}
    \label{fig:IMF}
\end{figure}

We run the simulation until star formation terminates due to feedback, which allows us to report the IMF relatively unambiguously i.e., without significant contamination by still-accreting protostars \citep[e.g.][]{bate2003}. In Figure \ref{fig:IMF} we plot the stellar IMF predicted by the simulation. For comparison we plot the \citet{chabrier_imf} IMF with the standard slope of -2.35, and a maximum-likelihood fit assuming the stellar masses are independently sampled,\footnote{Note that the IMF random sampling hypothesis is not necessarily an accurate description of how stars form in the simulation or in nature \citep{kroupa_2013_imf_review}, but we use it to perform our fit for lack of a more physically-motivated model for the correlations between the stellar masses, and because it is the most common assumption for fitting observations.}, which yields $\alpha=-2.0\pm0.1$. Note that the IMF predicted by Lagrangian simulations like ours, with finite mass resolution $\Delta m$, can only be reliable down to some multiple of $\Delta m$. This limit is generally assumed to be $\sim 50-100\Delta m$ \citep{bate_1997_resolution}. Here, we assume this IMF is incomplete below $\sim 100 \Delta m = 0.1 M_\odot$ due to numerical suppression of gravitational collapse for fragments smaller than this \citep[e.g.,][]{starforge.methods}, and focus on the portion of the IMF above this. Higher resolution is likely needed to comment on the abundance of brown dwarfs  \citep[e.g.,][]{bate:2009.hydro.sims,Offner_2009_radiative_sim}.

The  excess of massive stars found in previous versions of this simulation with isothermal MHD only \citepalias{guszejnov_isothermal_mhd} and with cooling and protostellar jets \citepalias{starforge_jets_imf} is now greatly suppressed, if not absent. The maximum stellar mass is $55 M_\odot$, and a total of 28 stars $>10M_\odot$ form in this $1560 M_\odot$ cluster, accounting for 35\% of the total mass. For comparison, at the time that the previous simulation with only protostellar jet feedback was halted (at a still-increasing SFE of 18\%), its most massive star was $460M_\odot$ and it had 41 stars $>10M_\odot$ containing 60\% of the total mass. This significant reduction in total and maximum mass of massive stars is due to feedback: the accretion of massive stars in the simulation is terminated by the creation of expanding feedback-driven bubbles due to some combination of radiative and wind feedback, as has long been theorized \citep{larson:1971.masslimit}. 

Overall, the predicted IMF is reasonably well-described by the \citet{chabrier_imf} form assuming $\alpha=-2$, with a slightly different shape in the range $0.2-1 M_\odot$. 
 The high-mass slope is more top-heavy than the commonly-adopted, ``canonical" value of -2.35 from \citet{salpeter_slope}, but it is well within the measured range for individual Galactic star clusters and OB associations \citep{massey_2003_imf_slope,kroupa_2013_imf_review}. Figure \ref{fig:fb_energy} also showed that the final specific bolometric luminosity $L_{\rm tot}$ and ionizing photon production rate $\mathcal{Q}$ are very close to those expected for a simple stellar population with a \citet{chabrier_imf} IMF, so it would difficult to distinguish the IMF of the simulated cluster from a canonical IMF by means of photometry \citep[e.g.][]{fumagalli_2011_universal_imf}.


Because the simulated IMF no longer has any feature in obvious, significant tension with present observations (as has also been found in various other works with similar physics, albeit with smaller cluster masses and statistics --  \citealt{Cunningham_2018_feedback,mathew:2021.imf.jets}), our programme of refining the star formation model on the basis of how well it reproduces the IMF will require more sensitive tests. In particular, it will be important to perform true mock observations and to compare with observational data using rigorous statistical methods. This is necessary to model the many biases and systematic effects that arise when attempting to measure the IMF in real systems \citep{kroupa_2013_imf_review,Hopkins_A_2018_IMF_obs_review}. A detailed presentation of IMF results from a broader suite of {\small STARFORGE} simulations with the full physics package will be discussed in an upcoming paper (Guszejnov et al., in prep.). 

\subsection{Chronology and duration of individual star formation}
\label{sec:tform}
\begin{figure}
    \centering
    \includegraphics[width=\columnwidth]{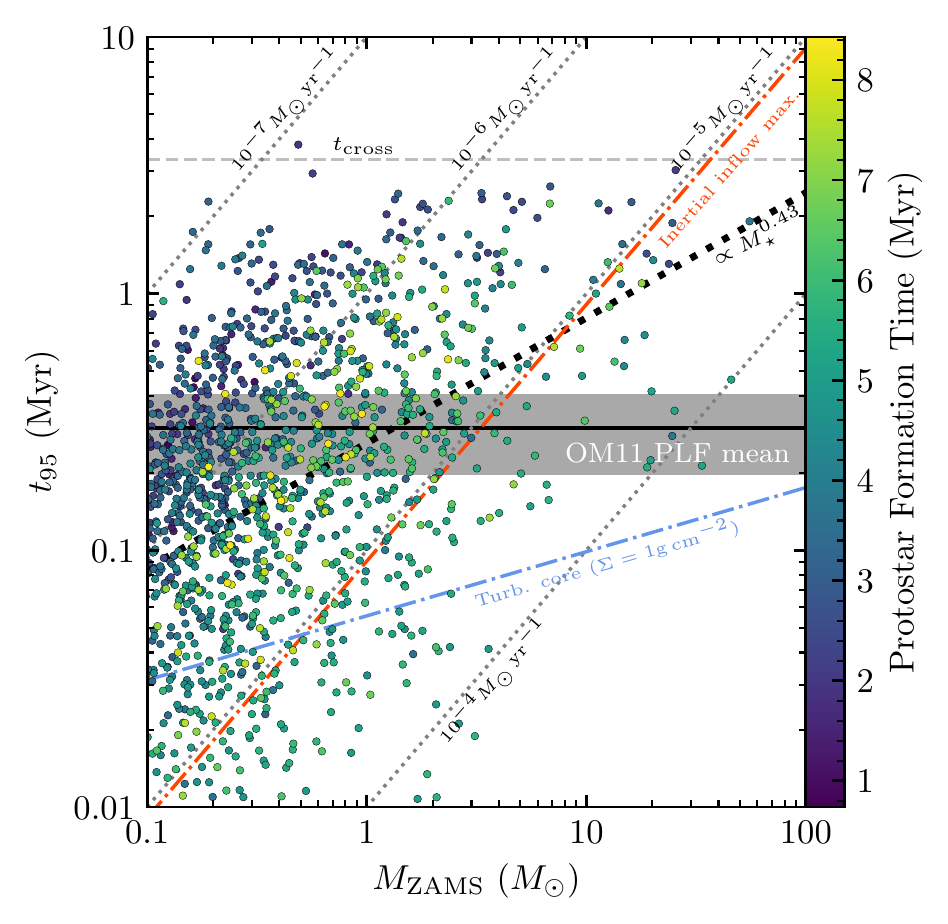}\vspace{-8mm}
    \caption{Time from protostar formation required to accrete 95\% of a star's mass, as a function of $M_{\rm ZAMS}$. Points are color-coded by the time that the protostar originally collapsed (i.e. the ``seed" formation time). Diagonal lines plot a range of average accretion rates $\dot{M} = 0.95M_{\rm ZAMS}/t_{\rm 95}$, including the maximum accretion rate predicted by the \citet{padoan_2019_massive_sf} inertial inflow model (orange dot-dashed). We also plot the GMC-scale turbulence crossing time $t_{\rm cross} = R_{\rm 0}/\sigma_{\rm 3D,0}=3.3 \rm Myr$ (grey dashed), the least-squares fitted relation $t_{\rm 95} \propto M_{\rm \star}^{0.43}$ (black dotted), the mean star formation timescale for low-mass stars inferred from the protostellar luminosity function in \citet{offner_mckee_2011_protostar_luminosity} (solid black with $\pm \sigma$ shaded region), and the formation timescale for massive stars for the fiducial turbulent-core model in \citet{Mckee_tan_2003_turbulent_core}, with $\Sigma=1\rm g\,cm^{-2}$ (blue dot-dashed).} 
    \label{fig:mstar_vs_dt}
\end{figure}

The time required for a star to assemble its mass can provide important clues about the formation mechanism \citep{offner_mckee_2011_protostar_luminosity}. In Figure \ref{fig:mstar_vs_dt} we plot the time required for a star to accrete $95\%$ of its eventual accreted mass, $t_{\rm 95}$, as a function of $M_{\rm ZAMS}$, as in \citet{Haugbolle_Padoan_isot_IMF} and \citet{ padoan_2019_massive_sf}. 

The average value of $t_{\rm 95}$ in our sample is $0.38\rm Myr$, in good agreement with the star formation timescale of $0.3 \pm 0.1 \rm Myr$ inferred from the protostellar luminosity function in low-mass star-forming regions \citep[][also plotted for comparison]{offner_mckee_2011_protostar_luminosity}. However our data has both a large scatter about this value, and a systematic trend toward longer accretion timescales for greater stellar masses. An unweighted logarithmic least-squares fit gives 
\begin{equation}
    t_{\rm 95} = 0.3 \rm Myr \left(\frac{M_{\rm ZAMS}}{M_\odot}\right)^{0.43},
\end{equation}
also plotted on Fig. \ref{fig:mstar_vs_dt}. Hence, on average, more-massive stars take longer to assemble. 
Indeed, $t_{\rm 95}$ for massive stars can be as long as $\sim 3 \rm Myr$, on the order of the freefall time $t_{\rm ff,0} \sim 3.7 \rm Myr$ or cloud-scale eddy crossing time $t_{\rm cross} = R_{\rm 0}/\sigma_{\rm 3D,0} \sim 3.3 \rm Myr$, which fits the upper envelope of $t_{\rm 95}$ values in general, as in \citealt{padoan_2019_massive_sf}. Protostars that eventually become massive do not form systematically earlier or later than others, but because they take longer to become massive, massive stars finish forming later ($\sim 1 \rm Myr$) than the average star. This has various interesting implications that we discuss further in \S\ref{sec:massive.sf.lag}. 

The recent massive star formation model by \citet{padoan_2019_massive_sf} aims to account for the gas assembly time through ``inertial inflows," which are coherent flows that accumulate gas in central hubs. They derived a maximum accretion rate for massive stars fed by inertial inflows in a supernova-driven turbulent medium, $\dot{M}_{\rm max}=2.8 c_{\rm s}^3/G \left(\mathcal{M}_{\rm 0}/10\right)^3 \alpha_{\rm turb}^{-1}$, where $c_{\rm s} \sim 0.2 \rm km\,s^{-1}$ in 10K molecular gas and $\mathcal{M}_{\rm 0}$ refers to the RMS turbulent Mach number on the driving scale. In our simulation the initial conditions are $\mathcal{M}_{\rm 0}=14$ and $\alpha_{\rm turb}=2$, giving $\dot{M}_{\rm max}^{\rm II}=1.1 \times 10^{-5} M_\odot\,\rm yr^{-1}$ (plotted as a red diagonal line on Fig. \ref{fig:mstar_vs_dt}). This an order of magnitude less than the maximum average accretion rate we find. Hence, we conclude that our simulation results are not well-described by the inertial inflow model as proposed, despite the qualitative agreement of the shape of our $t_{\rm 95}-M_{\rm ZAMS}$ diagram (Fig \ref{fig:mstar_vs_dt}). This discrepancy is likely due to the many differences in our respective simulation setups, but it is not presently clear which dominates this effect.

The relatively-long ($1\rm Myr+$) timescale for massive star formation also makes it impossible that most massive stars in the simulation draw their mass from dense ($\Sigma_{\rm gas} \sim 1 \rm g\,cm^{-2}$), gravitationally-bound turbulent cores \citep{Mckee_tan_2003_turbulent_core}. If this were so, then the stars would tend to accrete their mass on a timescale not much longer than the $0.1-0.3 \rm Myr$ freefall time of the core \citep{krumholz_2009_massive_sf,rosen_2016_massive_sf}. Figure \ref{fig:mstar_vs_dt} plots the prediction of the fiducial turbulent core model (assuming a gas surface density $\Sigma=1\rm g\,cm^{-2}$ and density profile $\rho \propto r^{-1.5}$) of \citet{Mckee_tan_2003_turbulent_core}, which lies below every massive star formed in the simulation.

A massive star formation scenario that is not obviously inconsistent with these results is competitive accretion \citep{zinnecker_1982_competitive_accretion, bonnell:2001.competitive.accretion,bonnell_2007_competitive_accretion_imf}. In this scenario, massive stars starting as intermediate-mass seeds accrete their gas from the GMC through gravitational capture in a manner reminiscent of Bondi-Hoyle-Lyttleton accretion ($\dot{M} \propto M^2$). Both low-mass and high-mass stars continue to accrete as long as sufficient gas is available, with their accretion rates being dictated by various factors. Numerical simulations without feedback have found this type of accretion to lead generically to an IMF slope $\alpha = -2$ \citep{Ballesteros-Paredes_2015_bhl,kuznetsova_2017_bhl, kuznetsova_2018_bhl}, similar to our result. However, these simulations did not include stellar feedback, and feedback is clearly responsible for limiting the maximum stellar masses in our simulation, so the overall scenario may be a combination of elements of competitive accretion and feedback regulation. 

The stellar accretion scenario can be characterized more definitively by analyzing accreta into the initially-bound core versus subsequently subsequently-captured components, determining how stellar accretion rates depend on physical properties, and determining the extent of the gas reservoirs feeding individual stars in space and time \citep[e.g.][]{smith_2009_cores, padoan_2019_massive_sf}. Our Lagrangian method makes this straightforward \citep[e.g.][]{bonnell_2007_competitive_accretion_imf}, and this analysis will be presented in an upcoming paper (Grudi\'{c} et al. in prep.).

\subsection{Stellar multiplicity}
\begin{figure}
    \centering
    \includegraphics[width=\columnwidth]{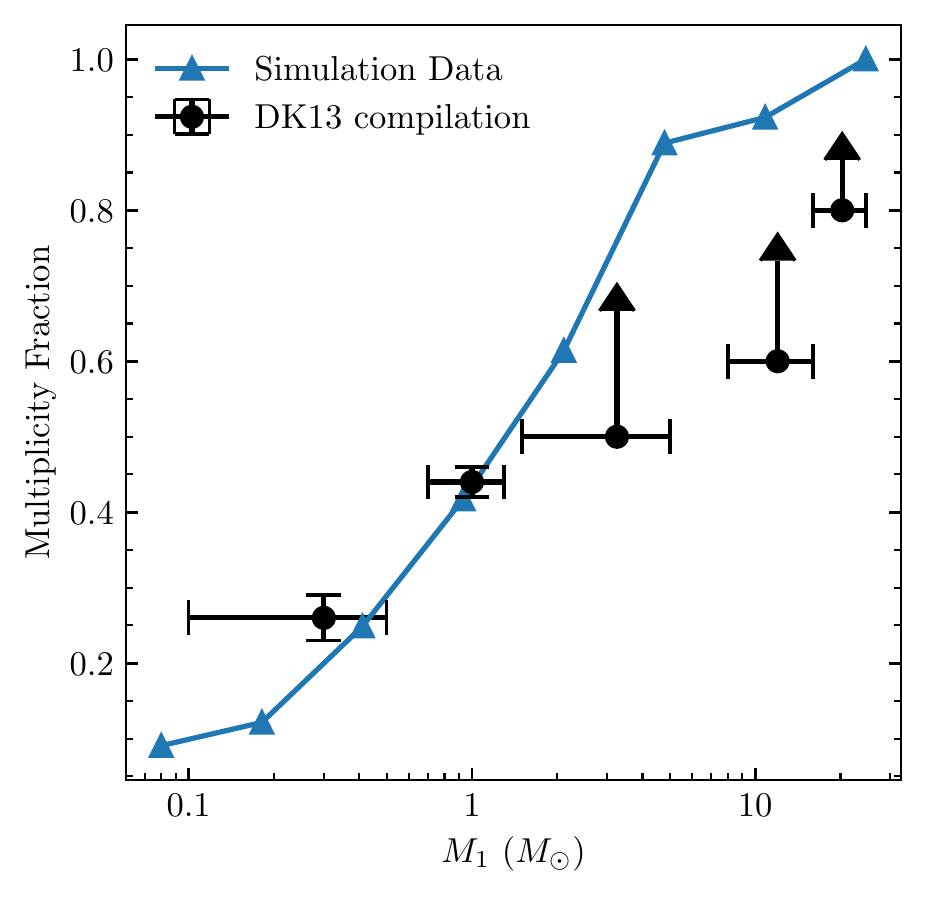}\vspace{-8mm}
    \caption{Fraction of stars in multiples at the end of the simulations, as a function of primary mass $M_{\rm 1}$, compared with the measurements compiled by \citet{duchene2013}. Upward arrows in the observational data indicate lower bounds.} 
    \label{fig:multiplicity}
\end{figure}

We identified bound multiple systems toward the end of star formation (at 8Myr) using the hierarchical grouping algorithm described in \citet[][]{bate:2009.hydro.sims}. Figure \ref{fig:multiplicity} shows the fraction of stars in bound multiple systems defined as
\begin{equation}
MF    = \frac{B+T+Q}{S+B+T+Q}
\end{equation}
for binaries, triples, or quadruples as a function of the primary mass of the system, $M_{\rm 1}$. We find this to be consistent with observations compiled by \citet{duchene2013}. Specifically, essentially all massive stars, roughly half of Solar-type stars, and relatively few ($\lesssim 25\%$) low-mass stars are in multiples at the end of the simulation. 

Many prior simulations have also obtained this result while considering more limited subsets of star formation physics \citep{bate:2009.hydro.sims,offner:2010.binary.turbfrag, guszejnov_correlation, Cunningham_2018_feedback}. However, our results show that, first, the trend extends to higher primary masses not attained in lower-mass cluster simulations, and second, the addition of feedback and other physics does not alter it significantly. The fact that simulations with such different physics obtained such similar multiplicity fractions -- both in agreement with observations -- indicates that other multiplicity statistics may be required to tease out the importance of different conditions and processes. 


The mass dependence of the multiplcity fraction is only one of many important statistics for characterizing stellar multiplicity \citep{moe.distefano:2017.multiples.review}. Additional statistics such as the multiplicity frequency and the mass-ratio and period distributions for this simulation and the extended {\small STARFORGE} suite, and their time dependence, will be presented in an upcoming paper (Guszejnov et al., in prep.).

\section{Discussion}
\label{sec:discussion}

\subsection{Comparison with previous GMC simulations}

To our knowledge the simulation presented here is the first to incorporate all major feedback mechanisms (winds, radiation, jets, and supernovae), so currently no other calculations are available that are directly comparable. However, many GMC simulations presented in the literature consider various subsets of the physics included here, so comparison with these may give clues about the effects of different physics and numerical details on star formation outcomes. This literature is extensive \citep{dale_feedback_review,krumholz_2019_cluster_review}, so we focus our discussion on simulations that start from similar GMC bulk properties and were run until star formation showed clear evidence of ending due to feedback, producing definite predictions for the outcome of SF. We perform this comparison with that general caveat that {\it none} of the simulations we compare with were initialized from precisely the same cloud microstate, so random variations due to the particular choice of initial state cannot be ruled out.



\subsubsection{Previous {\small GIZMO} simulations}
The {\small GIZMO} code was previously used to run GMC simulations with multi-physics cooling and heating, MHD, star formation, and feedback in the form of radiation, winds, and supernovae \citep{grudic_2016,grudic_2018_mwg_gmc,elephant, grudic_2020_cluster_formation}, but without self-consistent individual star formation or jets. Most directly comparable from these works are the GMC models with identical bulk properties ($M_{\rm 0}=2\times 10^4 M_\odot$, $R_{\rm 0}=10\rm pc$, and $\alpha_{\rm turb}=2$) simulated in \citet{grudic_2018_mwg_gmc}. These three simulations consistently found a final star formation efficiency $\epsilon_{\rm int} = 4\%$, half the $8\%$ of the current simulation (\S\ref{sec:sfe}), despite missing jet feedback. 

This may be due to a {\it nonlinear} feedback effect: without jets moderating star formation at early times, the \citet{grudic_2018_mwg_gmc} simulations formed stars much more rapidly at first (reaching peak $\epsilon_{\rm ff}$ at $<1 t_{\rm ff}$, vs. $1.5 t_{\rm ff}$ in Fig. \ref{fig:SFE}), meaning that significant feedback from other channels could emerge sooner, and disrupt the GMC from a less-collapsed phase. Feedback may have also been artificially enhanced by the star formation prescription: according to the \citet{su_imf_sampling} single-species O-star sampling scheme used in the previous simulations, the feedback rate reached that of a well-sampled IMF once $100-200 M_\odot$ had formed. In the current simulation, massive stars take significant time to form (>1 Myr), and the specific luminosity $L/M_{\rm \star}^{\rm tot}$ and ionizing flux $\mathcal{Q}$ only get close to their respective well-sampled IMF values when the cluster mass reaches $\sim 10^3 M_\odot$, 3 Myr after the start of SF (Fig. \ref{fig:fb_energy}). 



\subsubsection{{\small ATHENA} GMC simulations with UV feedback}

\citet{kim_2018_gmc_raytrace,kim:2020.gmc.raytrace} simulated a large suite of radiation hydrodynamical and MHD models of star-forming GMCs with the fixed-grid {\small ATHENA} code, with an initial setup very similar to ours. They relied on a sub-grid prescription for unresolved star formation but used adaptive ray-tracing \citep[e.g.,][]{wise.abel.adaptive.raytrace,rosen:2017.ray.tracing} to model feedback from ionizing and non-ionizing UV radiation, which is generally more accurate for single-scattered radiation than the M1 solver used here. The {\small M1e4R08} model from \citet{kim_2018_gmc_raytrace} with $M_{\rm 0}=10^4 M_\odot$ and $R_{\rm 0} = 8\rm pc$ is reasonably close to our model in parameter space, inviting comparison. Their model predicted $\epsilon_{\rm int} = 4\%$, again a factor of 2 lower than ours, so it is possible that the nonlinear effect of jet feedback and the time delay of massive SF may be important for setting the SFE. 


There is also some evidence that radiative feedback is driving photo-heated bubbles less efficiently in our simulation: in the \citet{kim_2018_gmc_raytrace} model the {\it photoevaporation} efficiency $\varepsilon_{\rm ion}$, the fraction of the cloud ionized by massive stars, was about 50\%, whereas in our model only $\sim 20\%$ of the cloud was ionized (Fig. \ref{fig:global.gas.props}), despite the higher SFE. One possibility is that our simulation captures a tendency for massive stars to form in denser environments, either due to suppression of fragmentation \citep{krumholz:2008.massive.sf.column.density} or due to more favorable conditions for accretion \citep{bonnell_2007_competitive_accretion_imf}. If so, those stars would irradiate denser gas and produce smaller HII regions, reducing the efficiency of ionization \citep{olivier_2020_hii}. Another possibility is that our higher resolution in dense regions allowed us to resolve the upper tail of the density PDF better, and thus more accurately model the formation of clumpy, porous gas structures that would be more resilient to ionization. However, we also cannot rule out the possible role played by the numerical method for radiative transfer: although our M1 solver simulates spherically-symmetric HII region expansion correctly \citepalias{starforge.methods}, it cannot represent the phase-space distribution of photons in more complicated geometries, which has little-explored implications for feedback in turbulent GMCs (see also \S\ref{sec:discussion.imfslope}).

\subsubsection{RAMSES-RT simulations with UV feedback}
\citet{geen_2017_sfe} performed a suite of adaptive mesh refinedment (AMR) radiation MHD simulations accounting for ionizing radiation with an M1 RT solver \citep{rosdahl:2013.ramses.rt}, with photons injected by sink particles with IMF-averaged ionizing fluxes (i.e. representing subclusters, rather than individual stars). Their model {\small L} with $M_{\rm 0}=10^4 M_\odot$, $R_{\rm 0}=7.65 \rm pc$, and $\alpha_{\rm turb}=1$ is most comparable to ours; this simulation found $\epsilon_{\rm int} = 4\%$. Here it seems especially plausible that the early onset of strong ionizing feedback leads to pronounced differences from our results: they note that the star formation history is punctuated by plateaus, due to star formation being terminated by feedback locally, and this occurs when as little as $100 M_\odot$ is in stars. At this cluster mass there is practically no ionizing feedback in our simulation (Figs. \ref{fig:global.star.props},\ref{fig:fb_energy}).

\citet{he_2019_gmc_fb} performed simulations with a similar setup to \citet{geen_2017_sfe}, and their models {\small S-F} and {\small M-F} are close in mass-radius space to ours (although their adopted initial $\alpha_{\rm turb}=0.4$ is significantly less than our $\alpha_{\rm turb}=2$) and obtained similar $\epsilon_{\rm int} \sim 4\%$ to \citet{geen_2017_sfe}. Unlike \citet{geen_2017_sfe}, \citet{he_2019_gmc_fb} argued their sink particle mass spectrum had sufficient physical significance to comment on the stellar IMF, and they invoked unresolved fragmentation to explain the discrepancy with the observed IMF. When assigning feedback rates to their sink particles, they assumed their cluster had the specific feedback rate of a well-sampled IMF and divided this feedback amongst their individual sink particles in a manner weighted according to the ionizing emission from a star of mass $M_{\star} = 0.3 M_{\rm sink}$. This model is incompatible with the picture in our simulation for two reasons. First, we find a fairly realistic IMF emerges naturally if multiple feedback mechanisms are accounted for (Fig \ref{fig:IMF}), without large corrections from unresolved fragmentation, and second, the assumption of an IMF-averaged mass-to-light ratio is not valid at early times. Their scheme for assigning sink luminosities is also inherently nonlocal (coupling the total cluster mass to individual sink feedback rates), so it is not clear that it should necessarily converge to a self-consistent picture of feedback on the scale of individual stars. 

Overall, the common element in our comparison with other simulations is that we find rather higher ($\times 2$) star formation efficiency than simulations that did not follow individual stellar formation and accretion self-consistently. This is plausibly explained by two features: first, our simulation accounts for jet feedback and the moderation of SF at early times, and second, it accounts for the finite time required for massive stars to form and thus for radiative feedback to become significant. Hence it is likely that the uncertain details of individual star formation and accretion are the leading source of uncertainty and discrepancy in GMC simulations \citep{elephant}. In future work it may yet be possible to account for such effects in lower-resolution simulations through sub-grid prescriptions calibrated to IMF-resolving simulations.

\subsection{The latency of massive star formation}
\label{sec:massive.sf.lag}
In our simulation, massive stars take systematically longer to form on average than lower-mass stars (\S\ref{sec:tform}, Figure \ref{fig:mstar_vs_dt}), and in particular, $>10 M_\odot$ stars finish accreting roughly 1Myr later than the average star. If massive stars do form with a characteristic time-lag  of one to a few Myr, there are several important implications for star formation. 

\subsubsection{Feedback timing}
 First, significant radiative, wind, and supernova feedback will emerge with a certain time-lag with respect to the onset of star formation in general. In observations, the latency of radiative feedback would affect diffuse emission diagnostics (e.g. free-free, H recombination lines, and IR), and stellar mass or formation rate measurements assuming coeval low- and high-mass star formation would generally underestimate the amount of star formation. In numerical models, delayed feedback can affect the dynamics of GMC evolution and disruption. Lower-resolution simulations using sub-grid star formation prescriptions do not generally account for such effects, except those that have investigated the importance of the delay explicitly \citep[e.g.,][]{keller_sne_uncertainty}. And indeed, from their numerical experiments \citet[][]{keller_sne_uncertainty} found that the specific choice of supernova delay can have important effects on the clustering of feedback and the overall galactic evolution. \citet{elephant} also found that SF prescriptions in which ionizing feedback emerged later led to systematically-higher final GMC-scale SFE, because the cloud was in a more advanced state of collapse by the time feedback switched on, and therefore, required more feedback to disrupt. 

\subsubsection{Photometric properties of young star clusters}
A lag in massive star formation on the order of Myr may also be important for the modeling of very young ($\lesssim 10 \rm Myr$) stellar populations in observations, e.g., for inferring star cluster properties such as mass and age from photometry in other galaxies \citep[e.g.][]{fall.chandar:2012.star.clusters}. Such measurements rely upon accurate model tracks in color space, which are typically generated assuming an ensemble of simple, coeval stellar populations sampled from an assumed IMF. If massive stars form with a systematic time delay, then a very young cluster will appear systematically dimmer and redder than a coeval stellar population of equal mass, leading to an overestimate of its age and an underestimate of its mass. 
In the opposite regime, once all massive stars have formed, the massive stars would be systematically younger than the average star in the cluster, and because they dominate the total flux the full population's age may be underestimated.

Photometric measurements of young star clusters are an important tool for constraining feedback, star cluster formation, and the IMF, and JWST will be used to study the earliest ($\lesssim 10 \rm Myr$) stages of cluster formation in this manner. Therefore it will be important to re-examine the working assumptions of photometric models of young stellar populations in light of our simulation results.

\subsubsection{The IMF in observations}
Lastly, the latency of massive star formation could have major implications for inferences about the nature of the IMF from observed young stellar clusters. Various systems are observed to have a deficit of massive stars compared to a standard IMF, even controlling for size-of-sample effects, e.g., Orion A versus the Trapezium Cluster \citep{hsu:2012.orion.a.lowmass.imf}. One interpretation is that different environments give rise to intrinsically-different stellar mass distributions. An alternative is that even if a proto-cluster displays a systematic deficiency in high-mass stars, it may still eventually form a cluster with a normal IMF, because the most massive stars haven't had time to accrete their eventual mass. In such systems the distinction between high- and low-mass star-forming regions would simply be one of evolutionary phase \citep[e.g.,][]{ghc}. 
If so, evolutionary phase is yet another factor to control for in IMF studies and would have to be characterized by an appropriate set of observables.



\subsection{The IMF slope, and simulation caveats}
\label{sec:discussion.imfslope}

The IMF predicted by this simulation, as well as various other simulations in the full-physics STARFORGE suite (Guszejnov et al, in prep.), has a high-mass slope $\alpha$ consistent with $-2$, shallower than the canonical \citet{salpeter_slope} value $\alpha=-2.35$. This means that the simulation IMF has a greater proportion of massive stars compared to commonly-assumed IMFs \citep[e.g.,][]{kroupa_imf,chabrier_imf}. Our result is not clearly ruled out by observations: the high-mass slope of the IMF remains the subject of ongoing investigation, complicated by the many practical difficulties and modeling uncertainties inherent in measuring the IMF in real galaxies and stellar systems \citep{imf_review,kroupa_2013_imf_review,offner_2014_imf_review,Hopkins_A_2018_IMF_obs_review}. Many individual clusters and associations in the Milky Way have indeed been reported to have $\alpha_{\rm 3} \gtrsim -2$, and compilations by \citet{kroupa_2013_imf_review} and \citet{weisz_2015_survey} found $\alpha=-2.36\pm 0.4$ and $-2.15 \pm 0.1$ respectively, not significantly incompatible with our $-2.0 \pm 0.1$.

However, clearly our slope is on the shallower end, so we also entertain the possibility that some limitation in the simulations is artificially enhancing the formation of massive stars, as was obviously true before we incorporated all feedback mechanisms \citepalias[e.g.][]{guszejnov_isothermal_mhd,starforge_jets_imf}. There are several possibilities for this.

\subsubsection{Jet physics}
     
     The simulation used the phenomenological jet feedback model of \citet{Cunningham_2011_outflow_sim}, with parameters $f_{\rm w}=f_{\rm K}=0.3$ based on observations, but the error-bars on the exact parameters are large, in part due to the difficulty of differentiating between directly-launched jet mass and entrained gas. For this reason, the {\it momentum} loading of jets $f_{\rm w} f_{\rm K}$ is better constrained than $f_{\rm W}$ and $f_{\rm K}$ individually. \citet{rosen_2020_jets_radiation} found their model, using parameters close to ours, was in good agreement with measurements of momentum injection rate (force) as a function of protostellar luminosity \citep{maud:2015.outflows,yang:2018.atlasgal.outflows}. However, if we adopted e.g. $f_{\rm w}=0.1$ and $f_{\rm K}=1$, the momentum would be similar but the jet velocities would approach $\sim 10^3 \,\rm km\,s^{-1}$ for $>10 M_\odot$ stars, which has been observed \citep{2010Sci...330.1209C}. Such jets would shock to temperatures $T \sim m_{\rm p} v^2 / k_{\rm B} >>10^6 \rm K$, above the peak of the atomic cooling curve, and hence would have an energy-conserving phase in which PdV work is done \citep{rosen_2021_winds}, enhancing feedback efficiency and potentially regulating massive SF. Our simulations with $f_{\rm W}=0.1$ and $f_{\rm K}=1$ in \citetalias{starforge_jets_imf} do show a hint of suppression of massive SF compared to the fiducial parameters, and we are following this up with full feedback physics.
     
\subsubsection{Unresolved disk fragmentation}
Features smaller than $\sim 100 \rm AU$ are not generally well-resolved in our simulation, and this prevents us from directly simulating the disks that are observed to form around protostars \citep[e.g.,][]{tobin:2020.orion.disks}. This prevents disk fragmentation on $\lesssim 100 \rm AU$ scales in the simulation, which might otherwise produce more-numerous, less-massive stars. This could potentially steepen the IMF if the effect is mass-dependent, which is plausible because high-mass disks are expected to fragment, whereas low-mass disks are more likely to be stable \citep{krumholz_2009_massive_sf,kratter10a}.  Assessing the implications of disk fragmentation for the IMF will require higher-resolution  ($\Delta x \sim 1 \rm AU$, $\Delta m \sim 10^{-5} M_\odot$) simulations. 
    
Our simulation also assumes ideal MHD, so even if we could resolve disks it is not clear that disks would survive magnetic braking \citep[e.g.][]{2003ApJ...599..363A}, and our preliminary experiments at higher resolution suggest not. However, the accuracy of the ideal MHD approximation breaks down at the low ionization fractions often found in protostellar envelopes and disks \citep{wurster:2021.nonideal}, so by neglecting non-ideal effects we may overestimate magnetic braking and prevent disk formation and fragmentation \citep{zhao_2020_nonideal_psd}.

\subsubsection{Feedback numerics and prescriptions}
The dominant feedback mechanism limiting the masses of the most massive stars is radiation, which we treat with a moments-based M1 solver (\S\ref{sec:radiation}, \citetalias{starforge.methods}). \citet{kim_2017_rhd_methods} performed a controlled comparison of GMC simulations accounting for single-scattered radiation pressure with an M1 solver \citep{skinner.ostriker:2013.m1,raskutti_2016} and an adaptive ray-tracing solver \citep{wise.abel.adaptive.raytrace, rosen:2017.ray.tracing}, and found the M1 solver systematically underestimated the effective strength of radiation pressure (overestimating the  final SFE compared to the ray-tracing run). Therefore we may also be underestimating the strength of radiative feedback, and consequently over-producing massive stars. 

We may also be underestimating feedback by neglecting post-main-sequence evolution (apart from our prescriptions for the Wolf-Rayet phase and supernovae, \S\ref{sec:methods}). The O stars that dominate the feedback budget in the simulation are expected to brighten (increasing by a factor of $\sim 2$ in bolometric luminosity), increasing the overall radiative feedback over time. This would not directly affect the dynamics of feedback in the immediate surroundings of a star as it accretes, but it might have an indirect effect, e.g. the brightening of stars throughout the cloud could cause it to disrupt earlier, cutting off the gas supply for massive stars.


\subsubsection{Lack of developed turbulence or driving}
GMCs are thought to be just one range of scales in a larger galactic turbulent cascade, wherein energy couples on scales on the order of the galactic scale height and cascades down to the dissipation scale \citep{maclow_star_formation_ism, excursion_set_ism,Padoan_2016_sne_driving}. For this reason, many simulations have modelled ongoing injection of turbulent energy on the largest scales of the cloud via a stirring procedure \citep[e.g.,][]{maclow:1999.turbulence}, instead of allowing turbulence to decay as we have. \citet{Haugbolle_Padoan_isot_IMF} and \citetalias{starforge_jets_imf} found that their respective runs with turbulence stirring in a periodic box obtained a \citet{salpeter_slope}-like IMF slope, even lacking feedback other than jets,\footnote{These were modelled implicitly in \citet{Haugbolle_Padoan_isot_IMF} via an assumed accretion efficiency factor $\epsilon_{\rm acc}$.} so it is possible that driving and/or fully-developed turbulence may be important missing ingredients. However, \cite{lane:2021.turbsphere} noted that stirring setups in the literature differ from ours in {\it multiple} regards apart from just the driving, and found that boundary conditions in particular can have significant effects on simulation results. The simulation setup developed in that work will be used to make a more controlled comparison of GMC simulations with full feedback physics, both with and without driving.


\section{Conclusions}
\label{sec:conclusion}

In this work we present some basic analyses of the first numerical simulation of a star-forming giant molecular cloud that simultaneously follows the formation, accretion, and feedback of individual stars, includes all major feedback channels (protostellar jets, radiation, winds, and SNe), and  evolves all the way to GMC disruption by feedback, producing a definite outcome of star formation. Our main findings are as follows:
\begin{itemize}
    \item The overall evolution of the GMC is qualitatively consistent with that anticipated by previous global GMC simulations without self-consistent star formation: the cloud collapses, forms stars at an accelerating rate, and is unbound and disrupted by feedback, which quenches star formation. This entire sequence takes $\sim 8 \rm Myr$, or roughly 2 global free-fall times (Figs. \ref{fig:renders},\ref{fig:global.gas.props}). 

    \item The star cluster assembles hierarchically from dense substructures, and the stars have systematic infall motions compared to the GMC as a whole (Fig. \ref{fig:global.star.props}). The star cluster has a brief compact phase with an effective radius $\lesssim 1 \rm pc$ but does not survive gas evacuation, so the cluster becomes an unbound association expanding at $2 \rm km\,s^{-1}$ with a ``Hubble-like" expansion law (Fig. \ref{fig:r_vs_vr}), similar to recent kinematics measurements enabled by {\it Gaia} \citep{kuhn:2019.gaia.cluster.kinematics}. 
    \item Of the different feedback mechanisms, protostellar jets are a dominant source of feedback momentum throughout most of the star formation process (Fig \ref{fig:fb_momentum}) and are important for regulating the IMF, but they cannot fully disrupt the cloud on their own \citepalias[see also][]{starforge_jets_imf}. Once sufficiently-massive stars form, their radiation and winds drive expanding bubbles that successfully disrupt the cloud and reduce the rate of star formation to a small fraction of its peak value. The one core-collapse supernova in the simulation occurs at 8.3Myr, too late to influence star formation significantly, but it does have a significant effect on the cloud kinematics (Fig. \ref{fig:global.gas.props}).
    
    \item We analyze various flavors of star formation efficiency in the simulation (\S\ref{sec:sfe}, Fig \ref{fig:SFE}), finding that the final integrated efficiency $\epsilon_{\rm int}=8\%$ is within the range estimated from statistical modeling of CO and $\rm H\alpha$ maps of nearby galaxies \citep{chevance_2020_gmcs}. The per-freefall efficiency $\epsilon_{\rm ff}$ behaves very dynamically, rising steeply to a crescendo of $18\%$, then dropping rapidly due to gas evacuation by feedback. We make some simple estimates of the observed SFE counterparts, and find good agreement with reported measurements based on tracing gas and stars with CO and free-free emission \citep{eve_lee_2016_GMC_sfe} and dust maps and YSO counts \citep{pokhrel:2020.gmc.sfe,pokhrel:2021.gmc.sfe}, respectively. These SFE proxies respectively over- and underestimate the instantaneous scatter of $\epsilon_{\rm ff}$ compared to the true value.
    
    \item Following the GMC evolution until termination of star formation by feedback allows us to measure a relatively unambiguous IMF. The IMF resembles the \citet{chabrier_imf} form with a high-mass slope $\alpha=-2\pm 0.1$ (Fig. \ref{fig:SFE}) and is significantly more realistic than previous iterations of this simulation without full feedback \citepalias{guszejnov_isothermal_mhd,starforge_jets_imf}. Radiation and/or winds from massive stars limit the maximum stellar mass in this cloud to $55 M_\odot$ (vs. $>400 M_\odot$ with jets only) and moderate the high-mass tail of the IMF overall. The integrated bolometric luminosity and ionizing photon rate of the cluster end up very close to that of an equal-mass cluster with a canonical IMF.
    
    \item We measure the time required for stars of different masses to accrete most of their mass after protostellar collapse (Fig \ref{fig:mstar_vs_dt}). The average value agrees with the $0.3 \pm 0.1 \rm Myr$ inferred from the protostellar luminosity function \citep{offner_mckee_2011_protostar_luminosity}, but the accretion timescale scales systematically with stellar mass, $\propto M_{\rm ZAMS}^{0.43}$ with significant scatter. Stars with $M_{\rm ZAMS}>10 M_\odot$  assemble their mass over $1 \rm Myr$ on average and can take as long as $3 \rm Myr$, suggesting they do not get most of their mass from dense, turbulent, bound cores (which would take an order of magnitude less time, \citealt{Mckee_tan_2003_turbulent_core}). Many stars also exceed the maximum accretion rate predicted by the inertial-inflow model \citep{padoan_2019_massive_sf} by an order of magnitude. The remaining major massive SF scenario -- competitive accretion -- is not ruled out in the simulation, but will require further analysis to characterize. The latency of massive star formation has various important implications for observational diagnostics and modeling of young star clusters (\S\ref{sec:massive.sf.lag}).

    
    \item We identify multiple star systems at the end of the simulation and measure the fraction of stars in multiples as a function of stellar mass, finding good agreement with observations (Fig. \ref{fig:multiplicity}). Unlike various other aspects of star formation, this quantity has also been reproduced by many prior calculations that considered more limited physics. Therefore other properties of stellar multiples are likely to be more-sensitive probes of star formation physics.
\end{itemize}

Overall, where key hallmarks of star formation are concerned, such as the IMF, star formation efficiency, stellar accretion, star cluster kinematics, and stellar multiplicity, we have reached the point where there is no longer any {\it blatantly} unphysical prediction from the model that must be fixed with additional physics, as was the case before various important mechanisms like magnetic fields and feedback were accounted for \citep{guszejnov_isothermal_collapse,guszejnov_isothermal_mhd,starforge_jets_imf}. On some level, the simulation successfully reproduces these key phenomena -- if and only if such physics is included. 

This represents some progress, but it would be premature to simply declare victory over the long-standing problem of modeling star formation.\footnote{The reader is directed to discussion of the various, yet-unresolved caveats of the {\small STARFORGE} framework: \S\ref{sec:discussion.imfslope} and \citetalias{starforge.methods}, \S5.2.} Rather, it is now time to look at the simulations and observations more closely.  Further progress on constraining star formation models will require more sensitive tests than the ones presented here, particularly by comparing observations with realistic mock-observations in a statistically-rigorous manner. It will also be useful to extend predictions down to smaller scales with increased resolution, e.g., to study protostellar disk and brown dwarf properties, which have their own constraints. In the mean time, we anticipate that the current setup will prove useful as a numerical laboratory for investigating SF physics in a controlled fashion, extrapolating predictions to environments beyond the Solar neighborhood (including the massive, $\sim 10^6 M_\odot$ complexes that dominate star formation), and interpreting ambiguous observations.


\section*{Acknowledgements}
 Support for MYG was provided by NASA through the NASA Hubble Fellowship grant \#HST-HF2-51479 awarded  by  the  Space  Telescope  Science  Institute,  which  is  operated  by  the   Association  of  Universities  for  Research  in  Astronomy,  Inc.,  for  NASA,  under  contract NAS5-26555. DG was supported by the Harlan J. Smith McDonald Observatory Postdoctoral Fellowship and the Cottrell Fellowships Award (\#27982) from the Research Corporation for Science Advancement. SSRO and ANR were supported by NSF CAREER grant AST-1748571. SSRO also acknowledges funding from NSF AST-1812747, NSF-AAG 2107942, NASA grant 80NSSC20K0507, and the Research Corporation for Science Advancement through a Cottrell Scholar Award (\#24400).
 CAFG was supported by NSF through grants AST-1715216, AST-2108230,  and CAREER award AST-1652522; by NASA through grant 17-ATP17-0067; by STScI through grant HST-AR-16124.001-A; and by the Research Corporation for Science Advancement through a Cottrell Scholar Award. ALR  acknowledges support from Harvard University through the ITC Post-doctoral Fellowship.  This work used computational resources provided by Frontera allocations AST20019 and AST21002, and additional resources provided by the University of Texas at Austin and the Texas Advanced Computing Center (TACC; http://www.tacc.utexas.edu). This research is part of the Frontera computing project at the Texas Advanced Computing Center. Frontera is made possible by National Science Foundation award OAC-1818253.

\section*{Data Availability}
The data and code supporting the plots within this article are available upon request to the corresponding authors. A public version of the {\small GIZMO} code is available at \url{http://www.tapir.caltech.edu/~phopkins/Site/GIZMO.html}. The complete sink particle dataset of the simulation, and data files containing the global quantities plotted in Figures \ref{fig:global.gas.props}, \ref{fig:global.star.props}, \ref{fig:fb_energy}, and \ref{fig:fb_momentum} are available at \url{https://github.com/mikegrudic/StarforgeFullPhysics}. 



\bibliographystyle{mnras}
\bibliography{bibliography} 








\bsp	
\label{lastpage}
\end{document}